\documentclass[secnumarabic,
amssymb, nobibnotes, nofootinbib, notitlepage, preprintnumbers
]{revtex4-1}
\usepackage{graphicx}

\usepackage{blindtext}	
\usepackage{amsmath,amssymb,amsthm}
\usepackage{latexsym,graphicx,color,subfigure}
\usepackage{enumerate}
\usepackage{color}
\usepackage{hyperref}
\usepackage{slashed}

\usepackage{amsmath,amssymb}
\usepackage{bm}
\usepackage{mathrsfs}
\usepackage{float}
\usepackage{subfigure}
\usepackage{braket}

\newcommand{\D}{\mathcal{D}}
\newcommand{\p}{\partial}
\newcommand{\Tr}{{\rm Tr}}

\newcommand{\bea}{\begin{eqnarray}}
\newcommand{\eea}{\end{eqnarray}}
\def\Tr{ \hbox{\rm Tr}}

\def\half{\frac{1}{2}}
\def\a{\alpha}

\def\half{\frac{1}{2}}

\def\l{\left}
\def\r{\right}

\def\Det{{\rm Det}}
\begin{document}
\title{
{Note on a solution to domain wall problem with the \\ Lazarides-Shafi mechanism 
in axion dark matter models }}
\author{Chandrasekar Chatterjee$^{1,2}$}
\email{chandra@phys-h.keio.ac.jp}
\author{ Tetsutaro Higaki$^{1,3}$}
\email{ thigaki@rk.phys.keio.ac.jp }
\author{ Muneto Nitta$^{1,2}$}
\email{nitta@phys-h.keio.ac.jp}
\affiliation{Research and Education Center for Natural Sciences,
Keio University, Yokohama 223-8521, Japan$^{1}$ \\ Department of Physics, Keio University, Yokohama 223-8521, Japan$^{2}$\\ Department of Physics, Keio University, Yokohama 223-8522, Japan$^{3}$
}

%

\begin{abstract}
Axion is a promising candidate of dark matter. After the Peccei-Quinn symmetry breaking, 
axion strings are formed and attached by domain walls when the temperature of the universe becomes comparable 
to the QCD scale. Such objects can cause cosmological disasters if they are long-lived. 
As a solution for it, the Lazarides-Shafi mechanism is often discussed through introduction of a new non-Abelian (gauge) symmetry.
We study this mechanism in detail and show configuration of strings and walls.
Even if Abelian axion strings 
with a domain wall number greater than one
are formed in the early universe, each of them is split into multiple Alice axion strings
due to a repulsive force between the Alice strings even without domain wall.
When domain walls are formed as the universe cools down, a single Alice string can be attached by a single wall 
because a vacuum is connected by a non-Abelian rotation without changing energy. 
Even if an Abelian axion string attached by domain walls
are created due to the Kibble Zurek mechanism at the chiral phase transition, 
such strings are also similarly
split into multiple Alice strings attached by walls in the presence of
the domain wall tension.
Such walls do not form stable networks since they collapse by the tension of the walls, emitting axions.
\end{abstract}
\maketitle

\newpage
\section{Introduction}
The standard model successfully explains various phenomena in experiments.
However there exist several unsolved problems.
One of the problems is the strong CP problem.
We may have a CP violating term in QCD: 
\begin{equation}
  \mathcal{L} = \theta \frac{g_s^2}{32\pi^2}G^{a\mu\nu}\tilde{G}^a_{\mu\nu},
\end{equation}
where $\theta$ is a constant parameter, $g_s$ is the strong coupling constant, 
$G^{a}_{\mu\nu}$ is the gluon field strength, and $\tilde{G}^{a}_{\mu\nu}$ is its dual. 
Measurements of the neutron electric dipole moment gives a constraint that $|\theta| \lesssim 10^{-10}$ \cite{Baker:2006ts},
while a naive expectation is that $|\theta| = {\cal O}(1)$. Without any mechanisms, this would require a fine-tuning.
The problem can be naturally solved
if one introduces the Peccei-Quinn (PQ) mechanism 
with a global $U(1)$ symmetry denoted by $U(1)_{\rm PQ}$
\cite{Peccei:1977hh, Peccei:1977ur, Weinberg:1977ma}. 
After the spontaneous $U(1)_{\rm PQ}$ breaking, 
a pseudo Nambu-Goldstone boson called the QCD axion appears.
When the $U(1)_{\rm PQ}$ symmetry is explicitly broken by QCD instanton effect
via chiral anomaly between $U(1)_{\rm PQ}$ and QCD, 
the axion vacuum is at a CP-conserving minimum of the potential and the axion solves the strong CP problem. 
The axion decay constant $f_a$ is of order 
the $U(1)_{\rm PQ}$ symmetry breaking scale, and the so-called axion window is given by
\begin{equation}
10^9 ~{\rm GeV} \lesssim f_a \lesssim 10^{12} ~{\rm GeV}.
\end{equation}
The lower bound comes from the SN 1987A neutrino burst duration \cite{Raffelt:2006cw}. 
The upper bound comes from the dark matter abundance by the misalignment mechanism
without a tuning of the initial misalignment, 
in which the coherent oscillation of the axion around the vacuum accounts for the abundance \cite{Preskill:1982cy,Abbott:1982af, Dine:1982ah}. 
See e.g. Refs.~\cite{Kim:2008hd, Kawasaki:2013ae} for reviews.

Physics of axion is related to the history of the universe.
Below temperature of $1$ GeV,
QCD instantons breaks the $U(1)_{\rm PQ}$ to $\mathbb{Z}_{N_{\rm DW}}$ symmetry, where $N_{\rm DW}$ is an integer called the domain wall number, then
there appear $N_{\rm DW}$ vacua.
Along with the PQ symmetry breaking, 
domain walls attached to strings
are formed 
\cite{Kibble:1982dd,
Vilenkin:1982ks,
Everett:1982nm,
Zeldovich:1974uw,
Preskill:1992ck} 
once one of the vacua is chosen.
The cosmological scenario depends on when the breakdown of the PQ symmetry takes place. 
If the PQ symmetry is broken before or during inflation, strings and walls 
are inflated away
because the axion field value becomes homogeneous over the scale of Hubble horizon
after inflation.
They cannot affect on
cosmological observations,
however,
there exist a stringent constraint on isocurvature perturbation produced by the axion
during the inflation \cite{Akrami:2018odb}. 
If the PQ symmetry is broken after inflation, 
such a constraint does not exist, but walls and strings
may survive until late time and can affect on evolution of the universe. We will focus on the latter case in this paper.

A stability of walls attached to strings is known to depend on $N_{\rm DW}$,
which is related to a topological charge.
For $N_{\rm DW}=1$, the domain wall attached to a string 
collapses owing to its tension, emitting axions.
On top of the misalignment, axions produced by their decays accounts for a fraction of the dark matter abundance 
\cite{Kawasaki:2014sqa,Klaer:2017ond}.
For $N_{\rm DW} > 1$, the domain walls attached to a string constitute complex networks, which are called string-domain wall networks.
The walls in the network pull each other with their tensions and
they do not shrink to a point, thus networks can be long-lived. 
They eventually dominate the energy density of the universe beyond those of radiation and matter, and conflict with the standard cosmology. 
This is called a domain wall problem \cite{Zeldovich:1974uw}.

For a solution to the domain wall problem, several ideas have been 
proposed so far \cite{Vilenkin:1981zs,Lazarides:1982tw,Kawasaki:2015lpf,Sato:2018nqy}. 
Among them, we will focus on an axion model
associated with a continuous non-Abelian gauge symmetry proposed 
by Lazarides-Shafi \cite{Lazarides:1982tw}\footnote{
A non-Abelian global symmetries are also viable for solving the domain wall problem
\cite{Kawasaki:2015ofa}.}.
One might think that a topologically stable domain wall is formed
when  $\mathbb{Z}_{N_{\rm DW}}$ 
symmetry is spontaneously broken by choosing the vacuum. 
However,
when the non-Abelian symmetry are also spontaneously broken 
at the same time and a combination of $\mathbb{Z}_{N_{\rm DW}}$ 
and the broken non-Abelian
rotation can make the vacuum invariant,
the vacua are also continuously connected also by the non-Abelian group 
without changing energy.
This is because the non-Abelian rotation is equivalent to a 
travel in the space of would-be NG modes.
Then, a topological property of such a domain wall for any $N_{\rm DW}$ becomes trivial
like in a case with $N_{\rm DW}=1$. Hence domain wall problem is solved.
This is called the Lazarides-Shafi mechanism.

However, behaviors of strings and walls in the mechanism 
has not been much discussed in literatures, 
while the authors of Ref.~\cite{Sato:2018nqy}
discussed the decay of Abelian axion strings to
Alice axion strings to solve the domain wall problem based on the mechanism.
In this model, 
we may have $N_{\rm DW}=2$ and a single Axion string is attached by 
two domain walls, seemingly having a domain wall problem. 
The Alice string produced by the decay plays a crucial role to realize a situation similar to the $N_{\rm DW}=1$ case,
in which one Alice string is attached by one domain wall. Hence the network is unstable in the presence of the domain wall tension.
(See Refs.~\cite{Abrikosov:1956sx,Nielsen:1973cs} for Abelian string
and also
Refs.~\cite{Hindmarsh:1994re,Vilenkin:2000jqa} for reviews of cosmic strings.)

Alice strings have a peculiar property that 
when the electric charge of a charged particle encircles around an Alice string, 
it changes its sign 
\cite{Schwarz:1982ec,Kiskis:1978ed}.
Other peculiar propeties such as
a topological obstruction, 
a non-local charge called the Chesire charge, 
and non-Abelian statistics,  
have been studied in the literature 
\cite{Alford:1990mk,Alford:1990ur,Alford:1992yx,
Preskill:1990bm,Bucher:1993jj,Bucher:1992bd,Lo:1993hp,Striet:2000bf}.
While a typical Alice string is present in 
an $SO(3)$ gauge theory with scalar fields in the fiveplet representation 
(a traceless symmetric tensor), 
recently it has been found that 
a $U(1) \times SO(3)$ gauge theory with 
complex triplet scalar fields  
also admits an Alice string,
which is a Bogomol'nyi-Prasad-Sommerfield state 
\cite{Bogomolny:1975de,Prasad:1975kr} and is stable, 
thereby possible to be embedded into supersymmetric theories 
\cite{Chatterjee:2017jsi,Chatterjee:2017hya}.
A global analog was known in the context of 
Bose-Einstein condensates in condensed matter physics 
\cite{Leonhardt:2000km,Ruostekoski:2003qx,Kobayashi:2011xb,Kawaguchi:2012ii}.
The Alice axion string proposed in Ref.~\cite{Sato:2018nqy} 
is the case that only the $U(1)$ part is global identified with axion,
while the $SU(2)$ part is a gauge symmetry. 

In this paper,
we show why the domain wall problem is solved physically in more detail,
focusing on dynamics of domain walls and two types of axion strings. 
It is found that even if Abelian axion strings are formed in the early universe, 
the string decays into multiple Alice axion strings owing to a repulsive force
between the Alice strings.
When domain walls are formed at the chiral phase transition
as the universe cools down, 
a single Alice axion string is attached by a single wall 
because a vacuum is connected by a non-Abelian rotation without changing energy. 
Such walls do not form stable networks since they collapse owing to the tension of the walls, emitting axions similarly to the $N_{\rm DW} =1$ case.
Also, at the chiral phase transition, two types of domain walls may be created by the Kibble-Zurek mechanism \cite{Kibble:1976sj,Zurek:1985qw},
and can be glued along an Abelian axion string.
The Abelian axion string is pulled by these domain walls and is splitted into 
multiple Alice axion strings, each of which is attached by one domain wall.
In either of these cases, the domain wall problem can be solved.

The rest of this paper is organized as follows. 
In Sec.\,\ref{review}, we briefly review the QCD axion and domain wall problem.
In Sec.\,\ref{model}, we introduce an axion model with heavy quarks
and a new gauge symmetry
for solving the domain wall problem. 
In Sec.\,\ref{CS}, properties of strings and domain walls are studied.
In Sec.\,\ref{DWCS}, we study domain walls attached to strings.
Sec.\,\ref{conclusion} is devoted to discussion and conclusions.


\section{Review of the QCD axion and domain wall problem}
\label{review}

In this section, we review the QCD axion based on
the Kim-Shifman-Vainstein-Zakharov
(KSVZ) model \cite{Kim:1979if,Shifman:1979if}
and domain wall problem for simplicity. 
(The Dine-Fischler-Srednicki-Zhitnitsky (DFSZ) model 
\cite{Dine:1981rt,Zhitnitsky:1980tq}
can be also discussed in a similar way.)
Let us consider a coupling
\begin{equation}
  \mathcal{L} = y\bar{Q}^i \Phi Q^i + {\rm h.c.}~~~(i= 1 ,2, \cdots , N_{\rm DW}).
\end{equation}
Here, $(\bar{Q},~ Q)$ are $N_{\rm DW}$ pairs of extra heavy quark in the $SU(3)_c$ gauge symmetry, and $\Phi$ is a complex scalar singlet under the Standard Model gauge symmetry. 
This coupling is invariant under the global $U(1)_{\rm PQ}$ symmetry:
\begin{align}
  &\bar Q \to e^{-i\theta}\bar Q~,~~Q \to Q~, ~~\Phi \to e^{i\theta}\Phi~.
\end{align}
Here $\theta$ is a transformation parameter.
Suppose that after inflation
$\Phi$ develops vacuum expectation value (VEV) $\eta$ and
the $U(1)_{\rm PQ}$ is spontaneously broken down.
Thus the scalar field is parametrized as
\begin{equation}
\Phi = \sqrt{2}\left(\eta + \frac{\sigma}{2}\right)\exp\left(i\frac{a}{2\eta}\right).
\end{equation}
Here $\sigma$ is supposed to be stabilized and we neglect it throughout this paper,
$a$ is the QCD axion. 
Note that a $U(1)_{\rm PQ}$ rotation of $a \to a + 2\pi \cdot (2\eta)$ 
is a symmetry.
All pairs of extra quarks obtain heavy mass $\sqrt{2}y \eta \times e^{i\frac{a}{2\eta}}$.
After integrating out extra quarks with the rotation of $\bar Q Q \to e^{-i\frac{a}{2\eta}} \bar Q Q$,
we obtain
\begin{equation}
 \mathcal{L} = \frac{a}{f_a} \frac{g_s^2}{32\pi^2}G^{a\mu\nu}\tilde{G}^a_{\mu\nu},~~~
{\rm where}~f_a \equiv \frac{2\eta}{N_{\rm DW}}
\end{equation}
via the chiral anomaly between $U(1)_{\rm PQ}$ and $SU(3)_c$.
Thus $\langle a \rangle = 0$ means the CP-conserving vacuum. 
After the chiral symmetry breaking in QCD, gluons and light quarks are 
integrated out, and the axion potential can be written as
\begin{equation}
V (a) \simeq m_\pi^2 f_\pi^2 \left( 1 - \cos \left(\frac{a}{f_a}\right) \right) 
\sim \cos \left(N_{\rm DW}\frac{a}{2\eta}\right) .
\end{equation}
Here, $m_\pi$ is the pion mass and $f_\pi$ is the pion decay constant.  
As desired,  $\langle a \rangle = 0$ is obtained in the vacuum and the strong CP problem is then solved.
The axion mass is given by
\begin{equation}
m_a \simeq \frac{m_\pi f_\pi}{f_a} \simeq 6 \times 10^{-4} {\rm eV} 
\bigg(\frac{10^{10}{\rm GeV}}{f_a} \bigg).
\end{equation}
Note that $a\to a + 2\pi f_a$
is the ${\mathbb Z}_{N_{\rm DW}}$ symmetry
against the above potential, in addition to the original larger
symmetry of $a \to a +2\pi N_{\rm DW}f_a $. Thus, 
there exist $N_{\rm DW}$ vacua.
Once one of the vacua is chosen,
${\mathbb Z}_{N_{\rm DW}}$ symmetry is spontaneously broken 
and topologically stable domain walls (attached to Abelian strings) 
appear between vacua for $N_{\rm DW} > 1$.
When one classically travels from a vacuum to the next one in the axion field space,
it is necessary to climb the potential energy. 
The walls pull each other with their tensions and
they do not shrink to a point, thus can be long-lived. 
The presence of stable domain walls conflict with the standard cosmology,
because they eventually dominate the energy density of the universe beyond 
those of radiation and matter.
It is verified in simulations that domain walls 
survive until late time for $N_{\rm DW} > 1$, 
while they decay for  $N_{\rm DW} = 1$ 
or in the presence of bias potential for $N_{\rm DW} > 1$ 
\cite{Kawasaki:2014sqa,Klaer:2017ond}.
For $N_{\rm DW} = 1$, the axion dark matter abundance
produced by decays of walls and strings is estimated as 
\begin{eqnarray}
\Omega_a h^2 \sim 10^{-2} \bigg( \frac{f_a}{10^{10}{\rm GeV}}\bigg)^{1.19}.
\label{Omegaa}
\end{eqnarray}

\section{The model with a non-Abelian gauge symmetry}
\label{model}
Following the Ref.~\cite{Sato:2018nqy}, we explain the model to implement the Lazarides-Shafi mechanism in the KSVZ case.
We shall start with the hidden $SU(2)_H$ gauge theory on top of the global 
$U(1)_{\rm PQ} $ symmetry. The Lagrangian is given as
\begin{eqnarray}
\mathcal{L} = - \frac{1}{2} \Tr F_{\mu\nu} F^{\mu\nu} +  \Tr|D_\mu\Phi|^2 
- V(\Phi)
+ ( y \bar{Q} \Phi Q + {\rm h.c.})
\label{LModel}
\end{eqnarray}
where $F_{\mu\nu}$ is the $SU(2)_H$ gauge field strength,
$\Phi$ is a complex adjoint scalar field and $(\bar{Q}, Q)$ are extra quarks charged 
also under both  $SU(2)_H$ and $U(1)_{\rm PQ} $.\footnote{
To avoid overproduction of massive particles at a high temperature,
$SU(2)_H$ doublet scalar field with TeV mass is introduced. 
As a consequence, extra quarks need to be charged under $U(1)_Y$.
Further, there may exist observational signal,
but we will focus only on configurations of strings and walls.
}
Their charge assignment against $(SU(3)_c, SU(2)_H, U(1)_{\rm PQ})$ is as follows:
$\Phi: (1,3,1),~\bar{Q}: (\bar{3},\bar{2},-1)$ and $Q:(3,2,0)$.
$SU(2)_H$ adjoint fields can be expanded with the $SU(2)$ generators $\tau^a = \half \sigma^a$ as $\Phi = \phi^a \tau^a$ and $A_\mu = A_\mu^a\tau^a$, and
the covariant derivative is defined as $D_\mu\Phi  = \p_\mu\Phi - i g [A_\mu, \Phi]$.
Here $g$ is the $SU(2)_H$ gauge coupling.
The potential for $\Phi$ is given as
\begin{eqnarray}
\label{potential}
V(\Phi) =  \frac{\lambda_1}{2} \l( \Tr (\Phi^\dagger\Phi) - 2 \eta^2\r)^2 +  \frac{\lambda_2}{2}\Tr \big([\Phi, \Phi^\dagger]^2\big).
\end{eqnarray}
This is an usual potential for complex adjoint scalar field that breaks $U(1)_{\rm PQ} \times SU(2)_H$ spontaneously. 
Later we will consider an explicit violation term for $U(1)_{\rm PQ}$,
which is relevant to axion mass and domain wall construction.
The ground state is given by the solution of the equations
\begin{eqnarray}
\Tr \Phi^\dagger\Phi = 2 \eta^2 , \qquad [\Phi, \Phi^\dagger] = 0.
\end{eqnarray}
The vacuum solution we may generally choose as
\begin{eqnarray}
\label{vac}
\langle \Phi \rangle = 2 \eta \tau^1= \eta \sigma^1.
\end{eqnarray}
The VEV breaks $U(1)_{\rm PQ}$ and $SU(2)_H$, 
then there exists the unbroken $U(1)_H$ gauge symmetry with a generator of $\tau^1$.
This vacuum is invariant under $O(2)$ rather than $SO(2)$.
The elements are given as
\begin{eqnarray}
\label{H}
H = \{(1, e^{i \alpha \tau^1}), (-1, i(c_1 \sigma^2 + c_2 \sigma^3)e^{i \alpha \tau^1})\}.
\end{eqnarray}
Here $\alpha, ~c_{1,2}$ are parameters and the conditon of 
$c_1^2 + c_2^2 =1$ is satisfied.
The first entry of each element of $H$ is $U(1)_{\rm PQ}$ 
and the second entry is the element of $U(1) \subset SU(2)_H$ group,
which can act as adjoint representation on $\langle \Phi \rangle$. The first element is usual element of $U(1)_H \subset O(2)$ connected to identity element $(1,1)$. However, the second element is a non-trivial $\mathbb{Z}_2$. Note that $e^{i\pi Q_{\rm PQ}} = -1$
for $Q_{\rm PQ}=1$, where $Q_{\rm PQ}$ is the PQ charge of the $\Phi$,
and $ e^{i\pi \tau^a} = i\sigma^a$ for $a=2,3$:
The first entry is $\pi$ rotation of $U(1)_{\rm PQ}$ and the part of second entry $i(c_1 \sigma^2 + c_2 \sigma^3)$ gives $\pi$ rotation of broken $SU(2)_H$ 
around an axis orthogonal to $\tau^1$.
So the second element of $H$ is referred to the disconnected elements of 
$O(2)$.

More specifically we may describe the symmetry
breaking procedure as
\begin{eqnarray}
G = U(1)_{\rm PQ} \times \frac{SU(2)_H}{\mathbb{Z}_2} \simeq U(1)_{\rm PQ} \times SO(3)_H \longrightarrow H = \mathbb{Z}_2 \ltimes U(1)_H \simeq O(2),
\end{eqnarray}
where $\ltimes$ denotes the semi-direct product because the $\mathbb{Z}_2 $ 
caused by $\pi$ rotation in the above discussion
does not commute with the unbroken $U(1)_H$
generated by $\tau^1$.
The vacuum manifold\footnote{
${\mathbb Z}_2$ in the original $SU(2)_H$ is the center acting on $\Phi$ trivially.
Even though there exist doublets, the vacuum manifold does not change unless they develop VEVs.  
} is found to be
\begin{eqnarray}
\frac{G}{H} = \frac{U(1)_{\rm PQ} \times SO(3)}{O(2)} = \frac{S^1\times S^2}{\mathbb{Z}_2}.
\end{eqnarray}
The fundamental group is $\pi_1(G/H) = \mathbb{Z}$, and this shows the existence of strings. 
It is noted that $\Phi$ has six degrees of freedom.
In this vacuum, two of three NG modes associated with the broken $SU(2)_H$ 
are eaten by the gauge fields $A^2$ and $A^3$, which
get massive. 
The rest (pseudo) NG mode relevant to $U(1)_{\rm PQ}$ is the QCD axion 
denoted by $a$.
The remaining three modes of $\Phi$ also become massive in the vacuum \cite{Sato:2018nqy}.

By integrating out extra heavy quarks with a mass matrix of
\begin{eqnarray}
\Phi = 2\eta \, e^{ia/2\eta}\tau^1, 
\end{eqnarray}
we have
\begin{eqnarray}
\mathcal{L} = N_{\rm DW}  \frac{a}{2\eta} \frac{g_s^2}{32\pi^2}G^{a\mu\nu}\tilde{G}^a_{\mu\nu} 
+ N_{\rm DW}' \frac{a}{2\eta}  \frac{g^2}{32\pi^2}F^{'\mu\nu}\tilde{F}'_{\mu\nu} .
\end{eqnarray}
Here, $N_{\rm DW} =2$ and $N_{\rm DW}' = 3$,
and $F^{'\mu\nu}$ is the unbroken $U(1)_H$ gauge field strength.
After the chiral symmetry breaking in QCD, 
the axion potential can be written as
\begin{equation}
V (a) \simeq m_\pi^2 f_\pi^2 \left( 1 - \cos \left(N_{\rm DW} \frac{a}{2\eta}\right) \right) \equiv m_a^2 f_a^2 
\left( 1 - \cos \left(\frac{a}{f_a}\right) \right) .
\label{Vaxion}
\end{equation}
Here, $m_a \simeq m_\pi f_\pi/f_a $ and $f_a = \frac{2\eta}{N_{\rm DW}}$.
To study domain wall later,
we parametrize this potential with $\Phi$ as
\begin{equation}
V_{\rm DW}(\Phi) =
\mu ( \Det \Phi + \Det \Phi^\dagger ) + {\rm const.}
\label{VDW}
\end{equation}
Here $\mu$ is assumed to be of ${\cal O}(m_a^2/N_{\rm DW}^2)$ constant\footnote{
$\mu$ actually depends on temperature of the universe via QCD instanton.
We will focus on a period during when domain walls are formed
below a temperature lower than $1$GeV
around which the axion mass is close to that at zero temperature.
}. A constant term is added to make the vacuum energy positive definite.
This form is motivated by the fact that axion and walls do not appear for $\Phi = 0$.
(See also \cite{Hiramatsu:2012sc}.)
$V_{\rm DW}(\Phi)$ will be added to $V(\Phi)$ of Eq.~(\ref{potential})
in the following sections 
about domain walls.
Since $N_{\rm DW}=2$ vacua are connected 
by $\mathbb{Z}_2$ embedded both in the $U(1)_{\rm PQ}$ and the broken $SU(2)_H$,
which acts as $\langle \Phi \rangle \to -\langle \Phi \rangle$, 
the domain wall problem is solved as seen later.
For a general $N_{\rm DW}$, a model with an $SU(N_{\rm DW})_H$ (gauge) symmetry 
is viable to implement the Lazarides-Shafi mechanism,
because ${\mathbb Z}_{N_{\rm DW}}$ of the center in $SU(N_{\rm DW})_H$
connects $N_{\rm DW}$ vacua.

Now, we have also monopole since $\pi_2(G/H) = \mathbb{Z}$.
It can be also a candidate of dark matter with mass of $10^{10}$ GeV
which is supposed to be
created by the first order phase transition of the $SU(2)_H$
at a high temperature around $3\times 10^9$ GeV \cite{Sato:2018nqy}.
The monopole can become a dyon 
with an electric $U(1)_H$ charge via the Witten effect \cite{Witten:1979ey}
when the axion cannot
cancel the CP phase in the hidden sector $\theta$-term.
Then the monopole may have mini electric charge via kinetic mixing 
between the photon and hidden photon in the $U(1)_H$.
We assume that the mini charge is sufficiently small to evade experimental bounds
\cite{DelNobile:2015bqo,Akerib:2016vxi}.
The monopole will not give further effects to solution to the domain wall problem
in the KSVZ case.
For $N_{\rm DW} >2$ with $SU(2)_H$, 
however, the Witten effect to the axion mass in the early universe \cite{Fischler:1983sc}
can give a significant effect to solve domain wall problem
in the presence of monopoles \cite{Kawasaki:2015lpf,Sato:2018nqy}.\footnote{
Monopole can also play an important role in suppressing isocurvature perturbation
of the axion
generated during inflation \cite{Kawasaki:2015lpf,Nomura:2015xil,Kawasaki:2017xwt},
if the PQ breaking occurs before/during the inflation.
}

\section{Construction of strings}
\label{CS}
Since the fundamental group of the vacuum manifold is an integer so we may expect topological string solutions.
In this section we discuss two kinds of string solutions.
One is an Abelian axion string originating from $U(1)_{\rm PQ}$ symmetry breaking
and this kind of strings can be thought of an usual global Abelian strings. 
Another is a non-trivial Alice axion strings which contain non-Abelian magnetic flux.  
This string can be classified by the elements of the unbroken group
$H = O(2)$. 
In the following, we discuss the splitting of one Abelian axion string 
into two Alice axion strings. 
We show full numerical results of the splitting 
and will set $V_{\rm DW}(\Phi) =0$ in this section.

\subsection{Abelian axion strings}
\begin{figure}[htbp]
\centering
\includegraphics[totalheight=4cm]{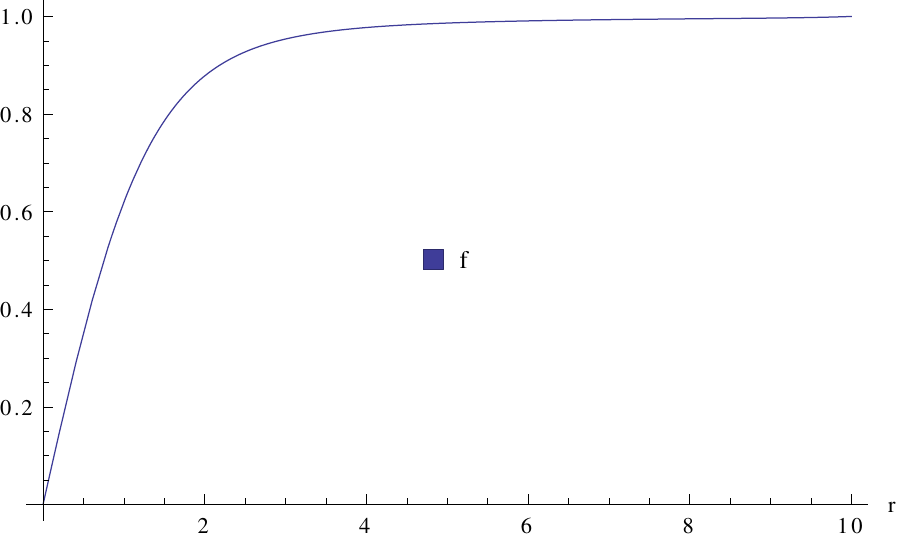}
\caption{Radial dependence of the profile function of $f(r)$ for $m =1, ~\eta = 2$ and $\lambda_1 = 0.2 $.}
\label{profileaxion}
\end{figure}

Here we would like to discuss the fully Abelian axion string which results from the $U(1)_{\rm PQ}$ breaking.
The axisymmetric ansatz for this string is given by
\begin{eqnarray}
\Phi = 2 \eta f(r)e^{i m \theta} \tau^1, \qquad A_\mu = 0
\end{eqnarray}
where $m$ is winding number, $r$ and $\theta$ are radial coordinate and azimuthal angle respectively. The boundary condition is given by $f(0) = 0, ~f(\rho) = 1$, 
where $\rho$ is the system size in the radial direction. The profile function $f(r)$ can be calculated from the axisymmetric equations of motion with this boundary condition. 
After inserting the above ansatz for $m=1$, we easily find the static Hamiltonian for 
$\Phi$ with Eq.~(\ref{LModel}). Two dimensional integral of the Hamiltonian density 
is given by
\begin{eqnarray}
\mathcal{H}^{(2)}
 = \int d^2x 2\eta^2\l\{ f'(r)^2 +\l( \frac{f(r)}{r}\r)^2 + \lambda_1 \eta^2 (f(r)^2 -1)^2\r\}.
\end{eqnarray}
Here prime denotes the derivative with respect to $r$.
Note that because the above Hamiltonian is written in a static case 
we have 
$\mathcal{H}^{(2)}  = -L$, 
where  $L = \int d^2x {\cal L}$ is 
two dimensional integral of the Lagrangian Eq.~(\ref{LModel}) with the ansatz.
With the radial coordinate, one dimensional Euler-Lagrange equation of the profile function reads
\begin{eqnarray}
- \frac{1}{r} \frac{d}{dr} r \frac{df(r)}{dr} + \frac{1}{r^2} f(r) + 2 \lambda_1 \eta^2 (f(r)^2 - 1) f(r) = 0.
\end{eqnarray}
The numerical result is shown in Fig.~\ref{profileaxion}, while
the analytical solution of the above equation is not known. 
%
%
The tension (energy per unit length) of these strings is found from the above 
$\mathcal{H}^{(2)}$. 
Approximate value of the tension far from the string core
is easy to compute if we insert the ansatz at a large distance
into the Hamiltonian: 
\begin{eqnarray}
T_{\rm PQ} =  
\mathcal{H}^{(2)}
 \simeq \int d^2x \Tr |\p_i\Phi|^2 \sim 2\pi \times 2 \eta^2 \times m^2 \log \rho.
\label{TAbel}
\end{eqnarray}
We may notice that energy is logarithmically divergent and energy depends on 
the square of winding number $m^2$.
The energy stored inside of the core is estimated as of order $\eta^2$, 
to which the scalar potential contributes,
and can be neglected at a large distance.
It is always energetically favorable to decay higher winding number string into lower winding strings. Since an Alice axion strings have $m=1/2$ as seen later, 
so an Abelian axion string with $m =1$ decays into two Alice axion strings.

\subsection{Alice axion strings}
The Alice axion string is a kind of topological string which changes the sign of electric charge of a probe particle with an original gauge symmetry
after one encirclement around the string. 
In our case, the generator of the unbroken $U(1)_H$ changes its sign 
with the $SU(2)_H$ after one rotation. This is because 
a particle charged under the $U(1)_H$ is affected by the broken $SU(2)_H$ 
flux inside the Alice string.
To understand this better, let us first consider
the field value rotating around an Alice axion string, which
depends on the azimuthal angle at a large distance $\rho$,
\begin{align}
\label{aliceaxion}
\Phi(\rho, \theta) & \sim \eta e^{i\frac{\theta}{2}} \left(
\begin{array}{ccc}
 0 &  e^{i\frac{\theta}{2}}     \\
  e^{-i\frac{\theta}{2}} & 0  \end{array}
\right) 
\equiv  e^{i\frac{\theta}{2}} \Omega(\theta) \Phi(\rho, 0) \Omega^{-1}(\theta), 
\nonumber
\\
A_{i} & \sim - \frac{1}{2 g} \frac{\epsilon_{ij }x^j}{\rho^2} \tau^3 ~(i,j=1,2),~
A_0 = A_3 = 0 .
\end{align}
Here, $x^2 +y^2 =r^2,~$
$\Phi(\rho, 0) = 2 \eta \tau^1=\langle \Phi \rangle$ and 
the holonomy $\Omega (\theta)$ 
rotating the $\Phi(\rho,0)$ by $\theta$ 
can be defined by the broken $SU(2)_H$ gauge field:
\begin{eqnarray}
\label{holonomy}
 \Omega(\theta) = Pe^{i g \int_0^\theta {\bf A\cdot dl}} = e^{i\frac{\theta}{2}\tau^3} 
\in SU(2)_H ,
\end{eqnarray}
where we used $\partial_i \theta =- {\epsilon_{ij }x^j}/{r^2} $.
It is easy to compute the (non-Abelian) flux in the broken $SU(2)_H$ 
trapped inside the string: 
\begin{eqnarray}
{\rm Flux} = \oint {\bf A\cdot dl} = \frac{\pi}{g} \tau^3
\end{eqnarray}

Ths $SU(2)$ holonomy can also be 
$\Omega(\theta) 
= e^{i\frac{\theta}{2}\tau^2}$ or more generally 
$\Omega(\theta) = e^{i\frac{\theta}{2}\hat n}$ 
with  
  $ \hat n = \sin\alpha \tau^2 + \cos\alpha \tau^3 = e^{i\alpha\tau^1} \tau^3 e^{- i\alpha\tau^1}$. 
  Correspondingly, the flux is along $ \hat n$. 
  The presence of $\alpha$ can be understood as follows.
  The unbroken symmetry $U(1)_H$ of the vacuum 
  is generated by $\tau_1$. 
  However, this $U(1)_H$ acts on the Alice string solution 
  with a parameter $\alpha$. 
  Namely, the Alice string configuration spontaneously breaks 
   $U(1)_H$ symmetry of the vacuum,
    implying the appearance of a Nambu-Goldstone mode in the vicinity of the Alice string. Therefore, $\alpha$ parametrizes a continuous family of the 
    Alice string solution of the same energy, and is a $U(1)$ modulus of the Alice string \cite{Chatterjee:2017hya}.

While $\langle \Phi \rangle$ in Eq.~(\ref{vac})  is invariant under the $U(1)_H$ with 
$\tau^1$,
$\Phi(\rho, \theta) $ in Eq.~(\ref{aliceaxion}) is not invariant under 
such unbroken elements in Eq.~(\ref{H})
because the unbroken transformation for $\Phi(\rho, \theta) $
becomes angle dependent. 
The $U(1)_H$ generator must be changed 
by the holonomies with the gauge field when it goes around 
the Alice string as
\begin{eqnarray}
Q_\theta = \Omega(\theta) Q_0 \Omega(\theta)^{-1}.
\end{eqnarray}
Here $Q_0 \propto \tau^1,$ 
thus transformations with $Q_{\theta}$ make $\Phi(\rho ,\theta)$ invariant
as 
$e^{i \beta Q_{\theta}} \Phi(\rho ,\theta) e^{-i \beta Q_{\theta}} = \Phi(\rho ,\theta) $,
where $\beta$ is a transformation parameter.
After encircling a full loop around the Alice string, we find that 
$\Omega(2\pi) = e^{i\pi\tau^3} \in \mathbb{Z}_2 \subset H$, 
hence the unbroken generator becomes
\begin{eqnarray}
\label{alicecharge}
Q_{2\pi} =  - Q_0.
\end{eqnarray}
This is nothing but the most characteristic property of the Alice string; 
the charge of a charged particle flips its sign when it encircles around an Alice string.

\begin{figure}[htbp]
\centering
\subfigure[\, ]{\includegraphics[totalheight=3.0cm]{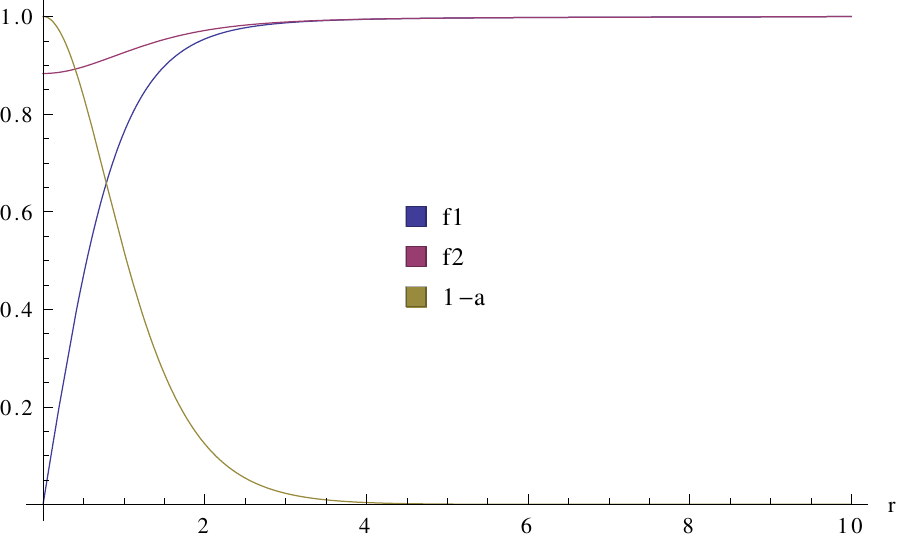}}
\subfigure[\, ]{\includegraphics[totalheight=3.0cm]{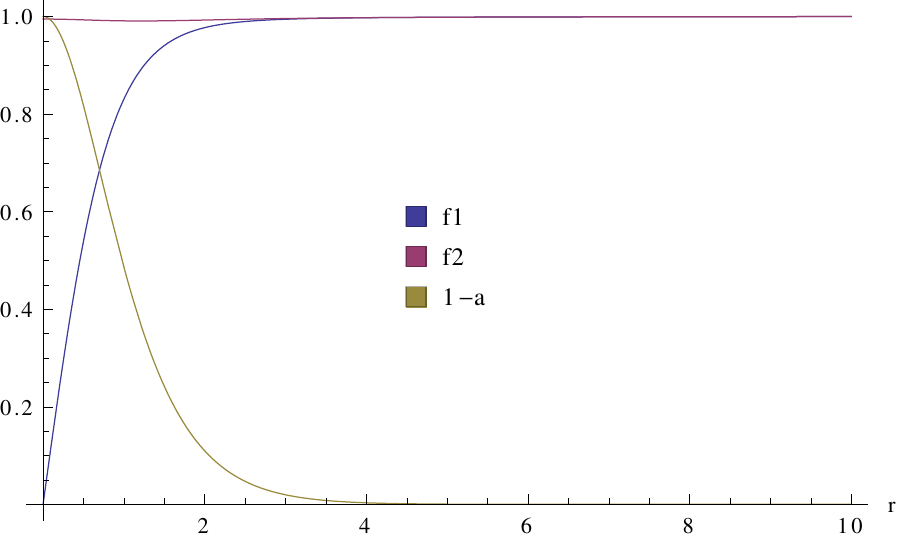}}
\subfigure[\, ]{\includegraphics[totalheight=3.0cm]{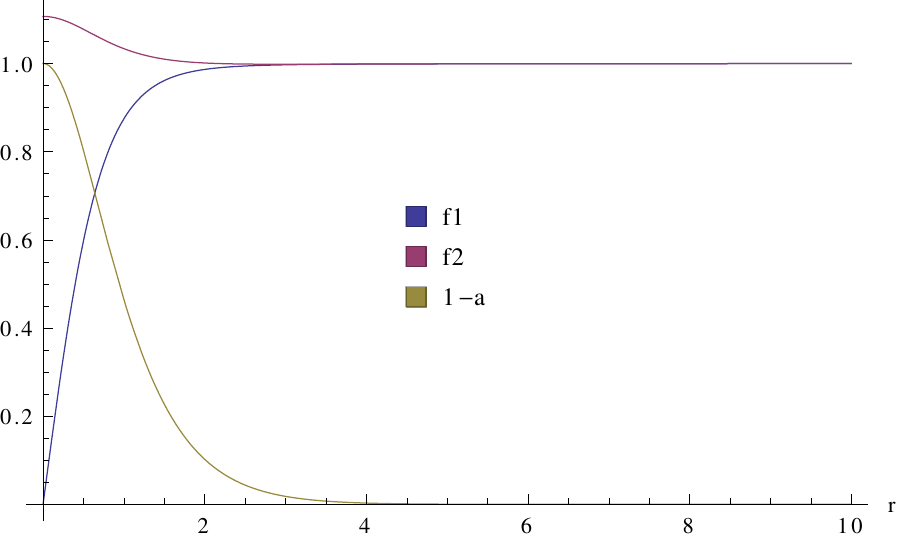}}
\caption{Radial dependence of the profile functions of $f_1(r),~f_2(r)$ and $a(r)$
for $g = 0.5$ and $\eta = 2$. 
The remaining parameters
are chosen as:
(a) $\lambda_1 = 0.2$ and $\lambda_2 = 0.2$ ($\frac{\lambda_1}{\lambda_2} =1  < 2$), 
(b) $\lambda_1 = 0.4$ and $\lambda_2 = 0.2$ ($\frac{\lambda_1}{\lambda_2} = 2$),
and
(c) $\lambda_1 = 0.6$ and $\lambda_2 = 0.2$ ($\frac{\lambda_1}{\lambda_2} = 3 > 2$.
This is used also in the Sec.\,\ref{DWCS}.}
\label{profilealice}
\end{figure}

To find a solution of the Alice axion string, 
we shall consider an axisymmetric ansatz as
\begin{eqnarray}
\Phi(r, \theta) = \eta 
\left(
\begin{array}{ccc}
0  & f_1(r) e^{i\theta}   \\
  f_2(r)&  0   
\end{array}
\right), 
\qquad
A_i = - \frac{1}{2 g} \frac{\epsilon_{ij }x^j}{r^2} a(r) \tau^3
~(i,j=1,2),~
A_0 = A_3 = 0 .
\end{eqnarray}
where $f_{1,2}(r)$ and $a(r)$ are profile functions of the scalar fields and gauge field with the boundary condition that
$f_1(0) = f_2(0) = 0, ~f_1(\rho) = f_2(\rho) = 1, ~a(0) = 0, ~a(\rho) = 1$. 
After inserting the above ansatz, 
two dimensional integration of the static Hamiltonian density
can be expressed as\footnote{
Although in the Hamiltonian there is no $1/g^2$ term in front of the gauge kinetic term,
ansatz of the gauge field includes $1/g$.
After all, we have the gauge coupling dependence only on the gauge kinetic term.
}
\begin{eqnarray}
\mathcal{H}^{(2)}
 &=& \int d^2x \l[ \frac{1}{2} \Tr F_{ij}^2 + + \Tr|D_i\Phi|^2 + V(\Phi)\r]\\
&=& 2\pi \int r dr \bigg[\eta^2\bigg\{ f_1'^2 + f_2'^2 + \frac{f_1^2}{r^2}\l(1 - \frac{a}{2}\r)^2 + \frac{f_2^2}{r^2} \frac{a^2}{4}
+ \frac{\eta^2\lambda_1}{2}(f_1^2+f_2^2 - 2)^2 + \eta^2\lambda_2(f_1^2 - f_2^2)^2\bigg\} 
+ \frac{1}{8 g^2} \frac{a'^2}{r^2}\bigg].
\end{eqnarray}
As in the Abelian string case, the equations of motion of the profile functions read
\begin{eqnarray}
&&
- \frac{1}{r} \frac{d}{dr} \l(r \frac{d}{dr}f_1(r)\r) + \frac{1}{r^2}\l(1 - \frac{a}{2}\r)^2 f_1(r) 
+  \lambda_1 \eta^2 (f_1(r)^2 + f_2(r)^2 - 2) f_1(r) 
+ 2 \eta^2\lambda_2(f_1^2 - f_2^2)f_1(r) = 0,
\\
&&
- \frac{1}{r} \frac{d}{dr} \l(r \frac{d}{dr}f_2(r)\r) + \frac{1}{r^2} \frac{a^2}{4} f_2(r) +  \lambda_1 \eta^2 (f_1(r)^2 + f_2(r)^2 - 2) f_2(r) - 2 \eta^2\lambda_2(f_1^2 - f_2^2)f_2(r) = 0,
\\
&&
- r \frac{d}{dr} \l(\frac{1}{r} \frac{d}{dr}a(r)\r) + 4 g^2 \eta^2 f_1(r)^2 \l(1 - \frac{a}{2}\r) + 2 g^2 \eta^2 f_2(r)^2 a(r) = 0.
\end{eqnarray}
The profile functions are solved numerically for several values of $\lambda_1$ and $\lambda_2$, and plotted in Fig.~\ref{profilealice}.
It is noted that in the equation of motion for $f_2$ 
the potential contribution can vanish
if $\lambda_1 = 2 \lambda_2$ is taken.
%
%

The tension far from the string core 
can be approximately computed with the ansatz at a large distance $\rho$ in Eq.~(\ref{aliceaxion}):
\begin{eqnarray}
T_{\rm Alice} = 
\mathcal{H}^{(2)}
 \simeq \int d^2x \Tr |\D_i\Phi|^2 \simeq
 \frac{\pi}{2} \times 2 \eta^2 \log \rho.
\end{eqnarray}
It should be noted that the energy of Alice axion string is also logarithmically divergent and 
the same as that of the Abelian axion string with $m =1/2$ in Eq.~(\ref{TAbel}). 
Here there would be contribution from $SU(2)$ magnetic field $\frac{1}{2} \Tr F_{ij}^2$,
however that we have neglected due to the logarithmic divergence of the leading term
and the energy stored inside the core of the Alice string is estimated as of order $\eta^2$, to which fluxes and the scalar potential contribute.
As mentioned in the previous subsection, 
an Abelian axion string (with $m =1$) can decay into two Alice axion strings 
(with $m =1/2$),
since the energy of the Abelian strings gets lower by the decay:
$1^2 > (1/2)^2 + (1/2)^2 = 1/2$ in the $m=1$ string tension unit, whereas
the winding number is conserved.
The remaining $1/2$ energy is thought to be converted to axions at the decay,
and they will contribute to a fraction of the final abundance of the axion.

%
\subsection{The decay of an Abelian axion string to two Alice axion strings}
In this subsection, we try to understand the decay of 
the Abelian axion string to the Alice axion strings.  
As seen above, the tension for the Abelian string  
is different from the Alice strings by a factor 4.
So the Abelian string (with $m=1$) is always energetically favorable 
to decay into two Alice strings (with $m=1/2$) with the conserved winding number. 
These two Alice axion strings produced by the decay 
must have opposite flux direction, since
the parent Abelian axion string contains no flux and
a field configuration at a large distance does not change through the decay.
The configuration of $\Phi$ at an angle $\theta$ and a large distance $\rho$,
which is far from the string core, can be in general written by
\begin{eqnarray}
\Phi(\rho, \theta) &=& h_0 h \langle \Phi \rangle h^\dagger, 
\quad \Omega = \{(h_0, h) | ~
h_0 
\in U(1)_{\rm PQ}, ~ h = P e^{ig\int^{\theta}_0 {\bf A\cdot dl}}  \in SU(2)_H \},\\
\Phi(\rho, 2\pi) &=& \Phi(\rho, 0) = \langle \Phi \rangle = 2\eta \tau^1,
\end{eqnarray}
where $\Omega = \Omega (\theta)$ 
is a path dependent rotation matrix with two entries around 
the strings in the axisymmetric case.
For an Abelian string, the gauge field is vanishing, whereas
for an Alice string with the positive flux 
the gauge field is given by the ansatz in the previous subsection.
For a full loop we have $\Omega(2\pi) =  (h_0(2\pi), ~h(2\pi)) \in H$ of $O(2)$, 
where $h(2\pi) = P e^{ig\oint {\bf A\cdot dl}}$.
For a rotation around the Abelian axion strings by $\theta$, 
$\Omega^0 (\theta) =(e^{i\theta},~1)$ and $\Omega^0(2\pi) =  (1, 1) \in H$, 
where zero denotes for one rotation around the Abelian axion string. 
For a rotation by $\theta$ around a single axisymmetric Alice axion string with 
a positive flux, $\Omega^+(\theta) = ( e^{i\frac{\theta}{2}},  e^{i \frac{\theta}{2}\tau^3})$ and $\Omega^+(2\pi) = \{-1, i \sigma^3\} \in H $. 
This is similar to $\Omega^-$ for the Alice string with a negative flux:
$\Omega^-(\theta) = ( e^{i\frac{\theta}{2}},  e^{-i \frac{\theta}{2}\tau^3})$ and $\Omega^-(2\pi) = \{-1, -i \sigma^3\} \in H $.

\begin{figure}[htbp]
\centering
\includegraphics[totalheight=4cm]{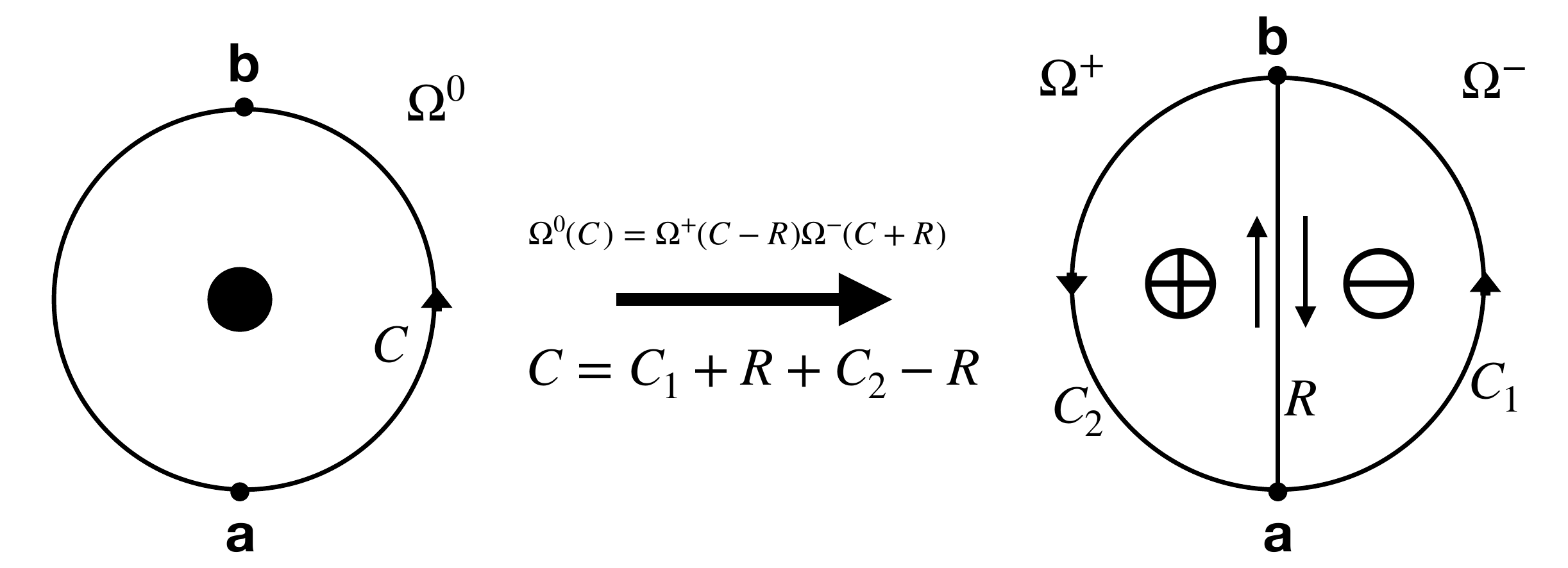}
\caption{Loops around the strings. Left: A loop $C$ around the Abelian axion string.
Right: A loop $C$ is decomposed to $C_1 +R$ and $C_2-R$,
where the former is a loop around the Alice axion string with the negative flux
and the latter is a loop around the Alice string with the positive flux.}
\label{SplittingL}
\end{figure}

Now let us understand what happens at the decay of the Abelian axion 
string to two Alice axion strings. 
Just at the time when a single Abelian axion string is split into two Alice strings, 
the boundary condition remains unchanged 
because it is understood with $\Omega$.
To show it,
we draw a loop $C$ around 
the Abelian axion string before splitting. In this case, the $\Omega$ is given by
$\Omega^0 (C)= (1,1)$.
Just after the splitting, we divide the loop in two parts as $C = C_1 + C_2$ and we close the loops by connecting the points $a$ and $b$ by the path $R$ as shown in the Fig.~\ref{SplittingL}  
In this case 
we have
\begin{eqnarray}
\Omega^{-}(C_1 +R) &=&  (h_0(C_1 +R) = e^{i \pi}, ~h(C_1 +R) = P e^{ig\oint_{C_1 +R} {\bf A\cdot dl}}) \\
\Omega^{+}(C_2 - R) &=&  (h_0(C_2 - R) = e^{i \pi}, ~h(C_2 - R) = P e^{ig\oint_{C_2 - R} {\bf A\cdot dl}}).
\end{eqnarray}
So we find 
\begin{eqnarray}
\Omega^0(C) = \Omega^{+}(C_2 - R)\Omega^{-}(C_1 +R) = 
(1, P e^{ig\oint_{C} {\bf A\cdot dl} }= 1).
\end{eqnarray}
This shows that field configurations at a large distances do not change,
while the splitting takes place.
%
\begin{figure}[htbp]
\centering
\includegraphics[totalheight=4cm]{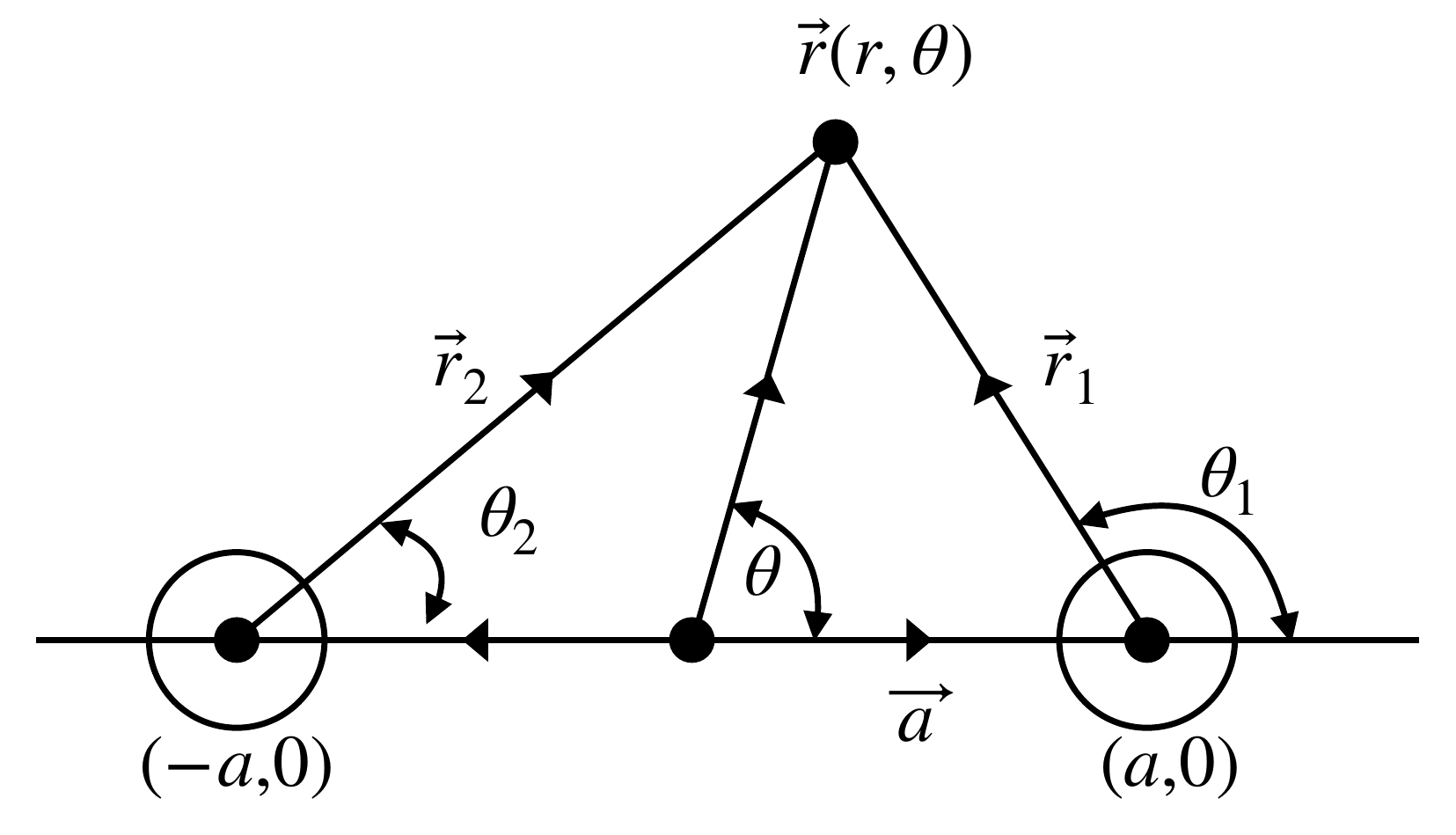}
\caption{A configuration of two Alice axion strings.
One Alice string with the negative flux is at $(a,0)$ in the $x$-$y$ plane,
whereas another with the positive flux is at $(-a,0)$. 
$|\vec r_{1,2}|$ are the distance vectors from the two Alice strings from the point 
${\vec r} =(r, \theta)$, and $\theta_{1,2}$ are the angles around them.}
\label{SplittingF}
\end{figure}
%

To estimate the force between two Alice axion strings, we take a large distance approximation, 
following Ref.~\cite{Nakano:2007dr} for the same problem 
in the context of non-Abelian strings in dense QCD 
\cite{Eto:2013hoa}. 
 Suppose that there exist two heavy Alice strings, which do not move,
on the $x$-axis in the $x-y$ plane at a large distance of $2a$ apart as shown in Fig.~\ref{SplittingF}. 
So the fields are approximately written as
\begin{eqnarray}
\Phi \sim e^{i (\theta_1 + \theta_2)} Pe^{ig \int_0^{\theta} {\bf {\bf A\cdot dl}}}\langle \Phi \rangle  Pe^{-ig \int_0^{\theta} {\bf A\cdot dl}},~
A_i(x, y) \sim A_i^{(+)}(x, y) +A_i^{(-)}(x, y)~(i=1,2),~
A_0= A_3 =0,
\end{eqnarray}
where $A_i^{(+)}$ ($A_i^{(-)}$) is the gauge fields relevant to 
the positive flux (negative flux) inside the Alice string 
in the absence of another string with the negative flux (positive flux).
So the energy at large distances can be approximately written by
\begin{eqnarray}
\mathcal{E}
\simeq \int d^2x \Tr |\D_i\Phi|^2
\simeq 2 \eta^2  \int d^2x \left[\p_i(\theta_1 + \theta_2)\right]^2 + \cdots
\end{eqnarray}
Here we neglected small contribution from the gauge field. 
It is found also that $\int d^2x \left[(\p_i\theta_1)^2+(\p_i\theta_2)^2\right]
 \simeq 2\times T_{\rm Alice} $.
So the interaction energy can be expressed as
\begin{eqnarray}
\mathcal{E}_{int} &&\simeq 4 \eta^2  \int d^2x \p_i\theta_1\p_i \theta_2 + \cdots \nonumber \\
&&= 4 \eta^2  \cdot 2\pi \int dr r \l[ \frac{r^2 - a^2}{r^4 + a^4 - 2r^2 a^2 \cos2\theta}\r] \nonumber \\
&& \simeq 4 \pi \eta^2 \log\l(\frac{\rho^2 + a^2}{a^2}\r).
\end{eqnarray}
Here we used $\partial_i \theta = -\epsilon_{ij}x^j/r^2$ and this is similar to
$\theta_{1,2}$.
The force between two Alice axion strings is found to be repulsive:
\begin{eqnarray}
\text{Force} = - \half \frac{\p \mathcal{E}_{int} }{\p a} 
\simeq + \frac{4 \pi \eta^2}{a} ~~~
{\rm for}~\rho \rightarrow \infty.
\end{eqnarray}
This repulsive force is mediated by the light QCD axion at a large distance apart.\footnote{
There exists an attractive force mediated by the massive gauge field 
between two Alice strings at a short distance.
It is expected, however, that such Abelian strings 
tend to decay into Alice strings in the presence of perturbations in the universe.
Even if Abelian axion strings survive until the chiral phase transition,
they are attached by domain walls
and can be split into Alice strings owing to the domain wall tension
as shown in Sec.\,\ref{DWCS}.
}
This is analogous to a Coulomb force between particles with the same charge
in two spatial dimension.
Hence, the distance between two Alice axion strings would increase with time and it is confirmed from numerical calculation. 
However, note that our simulation is done in a relaxation method but not 
in a real time dynamics.
%
\begin{figure}[htbp]
\centering
\subfigure[\, ]{\includegraphics[totalheight=3.3cm]{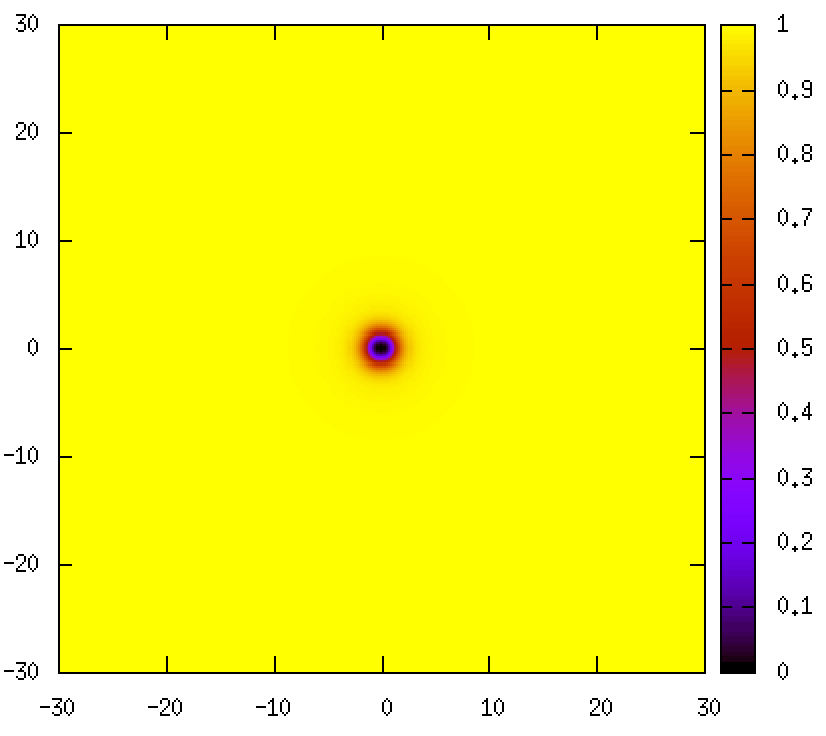}}
\subfigure[\, ]{\includegraphics[totalheight=3.3cm]{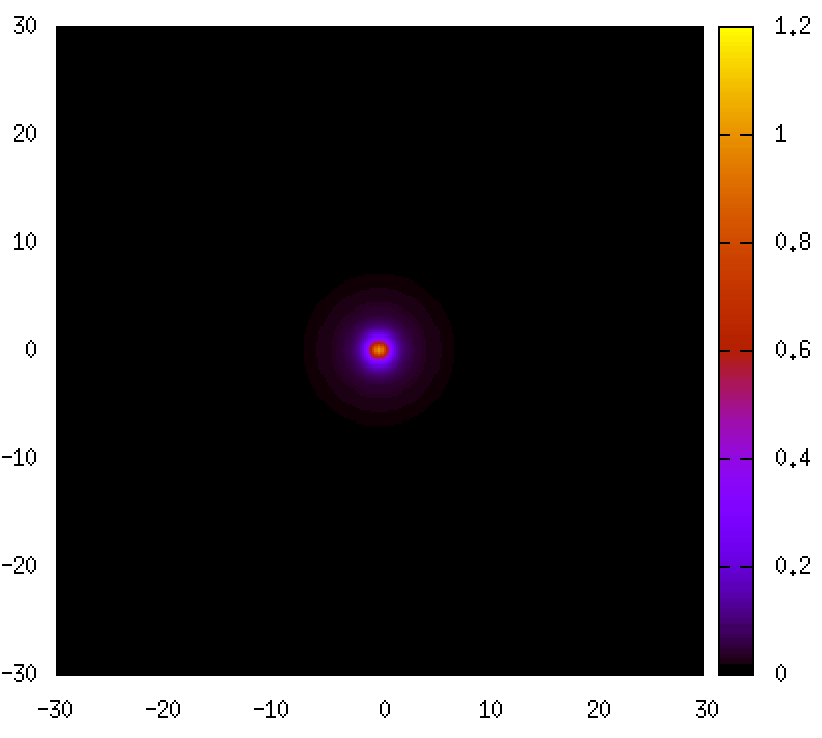}}
\subfigure[\, ]{\includegraphics[totalheight=3.3cm]{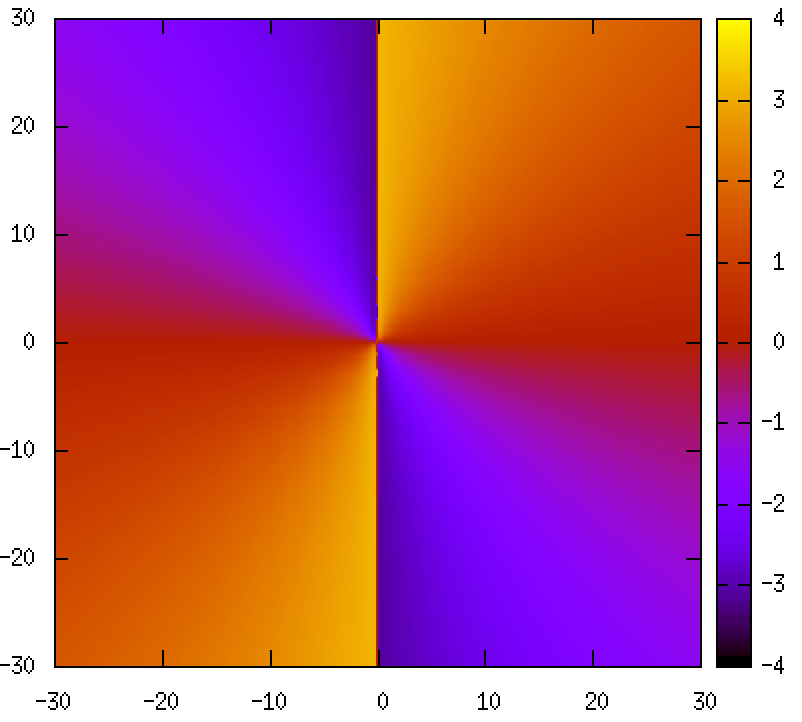}}
\subfigure[\, ]{\includegraphics[totalheight=3.3cm]{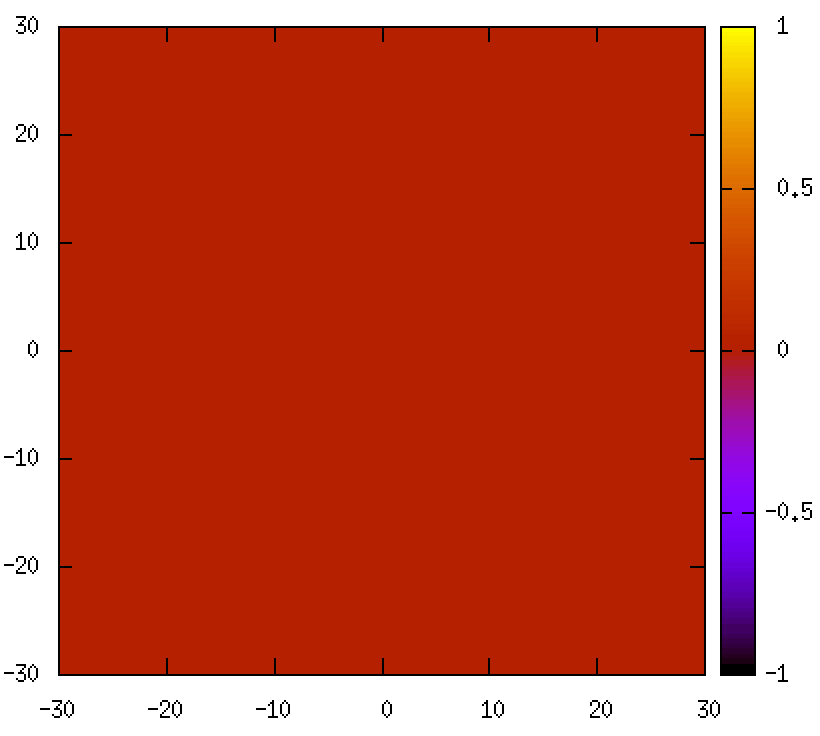}}
\caption{Figures of an Abelian axion string before the splitting
for $g=1,~\eta = 0.5, ~\lambda_1 = 2.0$ and $\lambda_2 = 0.25 $.
They show $2 \Tr(({\Phi}^2 )^\dagger\Phi^2)$,
the static energy,
the phase $ \Im\log\Tr\Phi^2(x,y)$ and 
non-Abelian magnetic field $F_{12}^3(x,y)$ (with a gauge fixing)
from left to right.
It is easy to check there is a phase jump by $2 \pi$ around each string solution.}
\label{Splitting1}
\end{figure}
\begin{figure}[htbp]
\subfigure[\, ]{\includegraphics[totalheight=3.3cm]{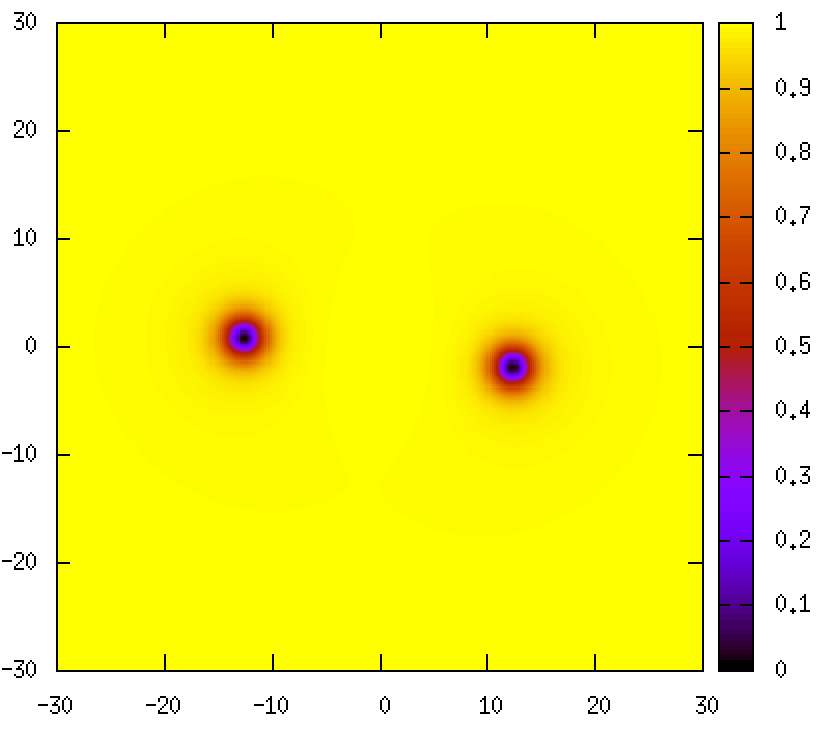}}
\subfigure[\, ]{\includegraphics[totalheight=3.3cm]{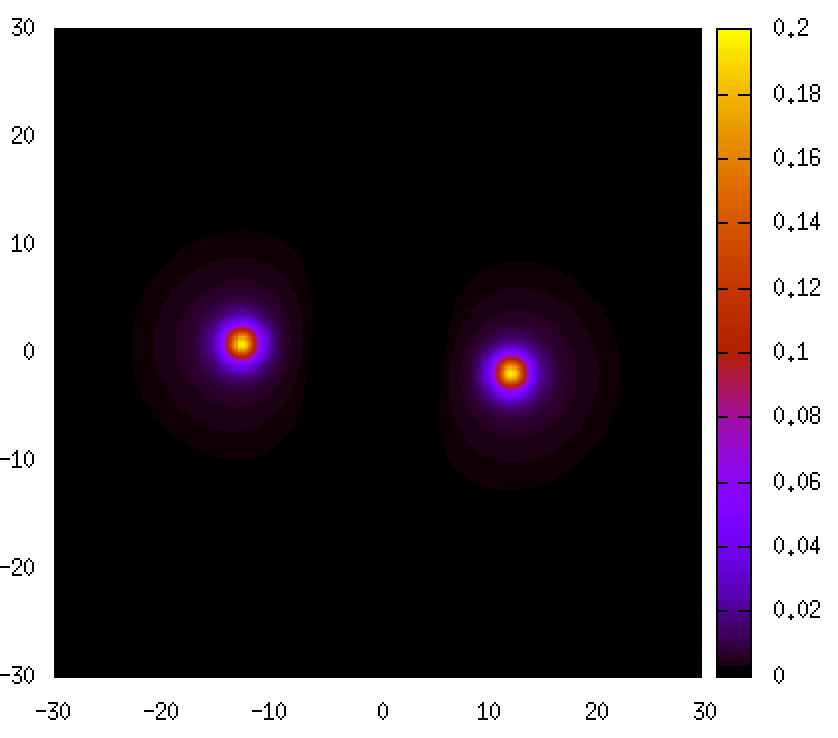}}
\subfigure[\, ]{\includegraphics[totalheight=3.3cm]{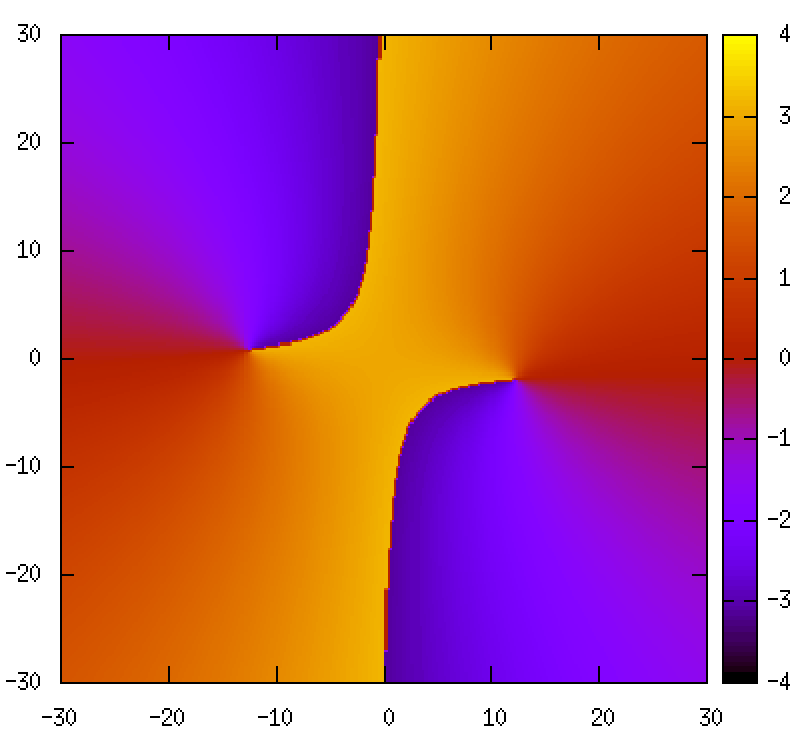}}
\subfigure[\, ]{\includegraphics[totalheight=3.3cm]{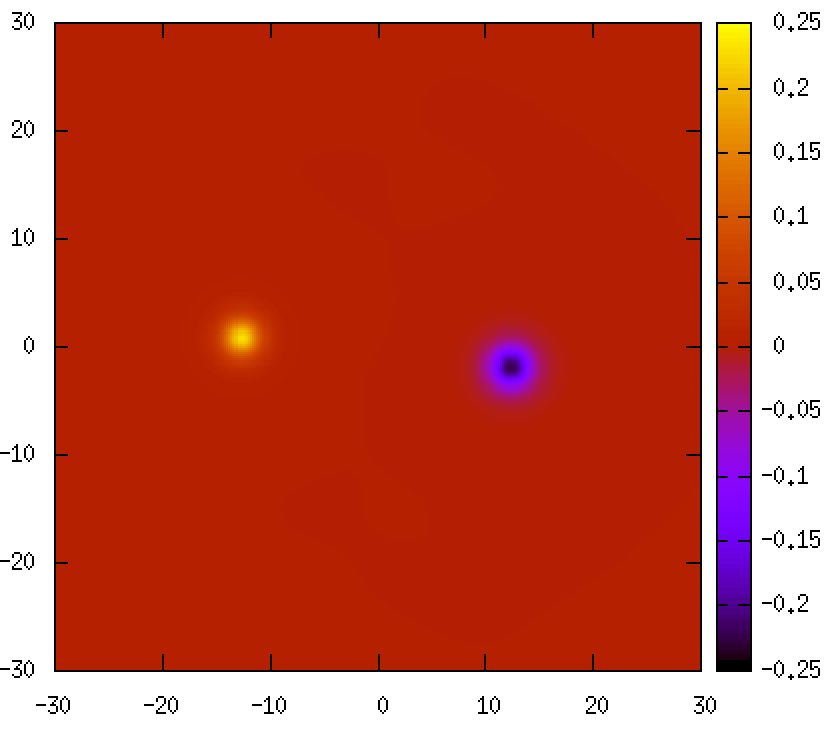}}
\caption{Similar figures of two Alice axion strings after the decay from the parent Abelian axion string.
}
\label{Splitting2}
\end{figure}

For numerical solutions, we used  $500 \times 500$ lattice 
with lattice spacing $0.2$. 
We relaxed the system in $1000000$ time steps with each time step 
$\Delta t = 10^{-3}$. 
The other parameters are taken as $g=1,~\eta = 0.5, ~\lambda_1 = 2.0$ 
and $\lambda_2 = 0.25 $.

\section{Domain wall-string composite}
\label{DWCS}
So far we discussed two types of axion strings and decay of a Abelian axion string to two Alice axion strings. 
In this section, we study 
configuration of
domain walls attached to the Abelian axion string or to the Alice string.
The former situation involving the Abelian strings may be realized
at the chiral phase transition through the Kibble-Zurek mechanism \cite{Kibble:1976sj,Zurek:1985qw}:
At the chiral phase transition, two kinds of domain walls may be created elsewhere.
If these domain walls collide, 
they can be glued along an Abelian axion string.
The latter situation involving the Alice strings will always take place in our case.
To find the domain walls attached to an Abelian axion string,
we start with $\mu \ne 0$ in the potential of Eq.~(\ref{VDW}).
The static Hamiltonian density (in $A_0 = 0$ gauge) is given by
\begin{eqnarray}
\label{HamiltonianDW}
 \mathcal{H} &=&   
 \frac{1}{2} \Tr F_{ij}^2 + + \Tr|D_i\Phi|^2 + \frac{\lambda_1}{2} \l( \Tr \Phi^\dagger\Phi - 2 \eta^2\r)^2 +  \frac{\lambda_2}{2}\Tr \big([\Phi, \Phi^\dagger]^2\big)
\nonumber
\\
&&+  
\mu\l(
\Det \Phi + \Det \Phi^\dagger \r) +{\rm const.}
\end{eqnarray}
We shall first discuss vacua in the presence of Abelian axion strings and 
the Alice axion strings and focus a parameter region, in which
$2\eta^2 \lambda_1 > \mu$ is satisfied, 
for $V_{\rm DW}$ not to affect $\langle \Phi \rangle$ significantly. 
This is natural for axion domain wall
since $\mu$ is expected to be of order $m_a^2$.

%
\subsection{Abelian axion string-domain wall composite}
\begin{figure}[htbp]
\centering
\subfigure[\, ]{\includegraphics[totalheight=4.0cm]{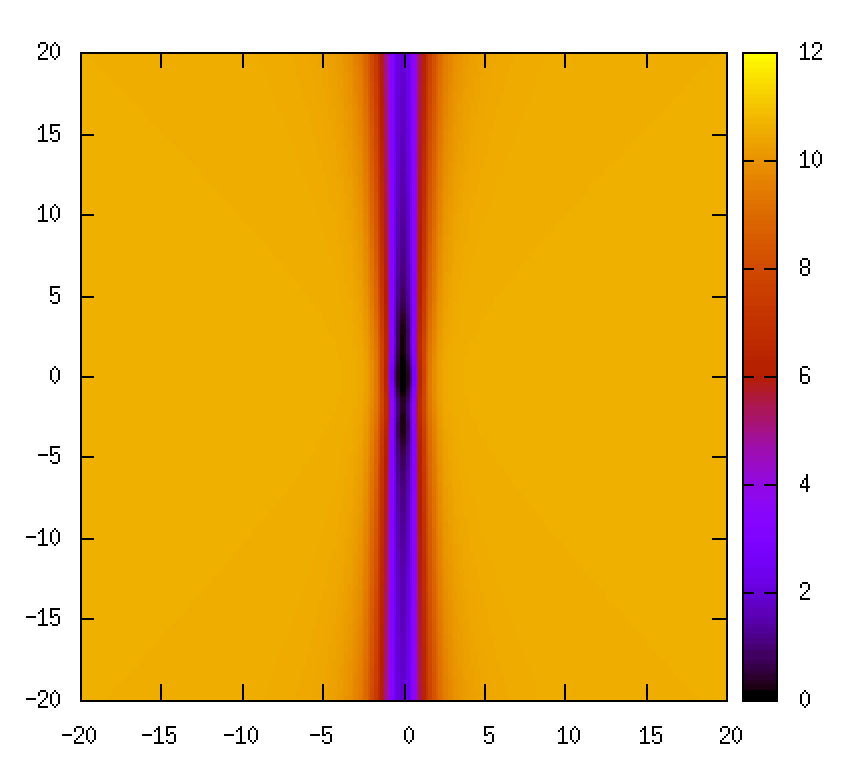}}
\subfigure[\, ]{\includegraphics[totalheight=4.0cm]{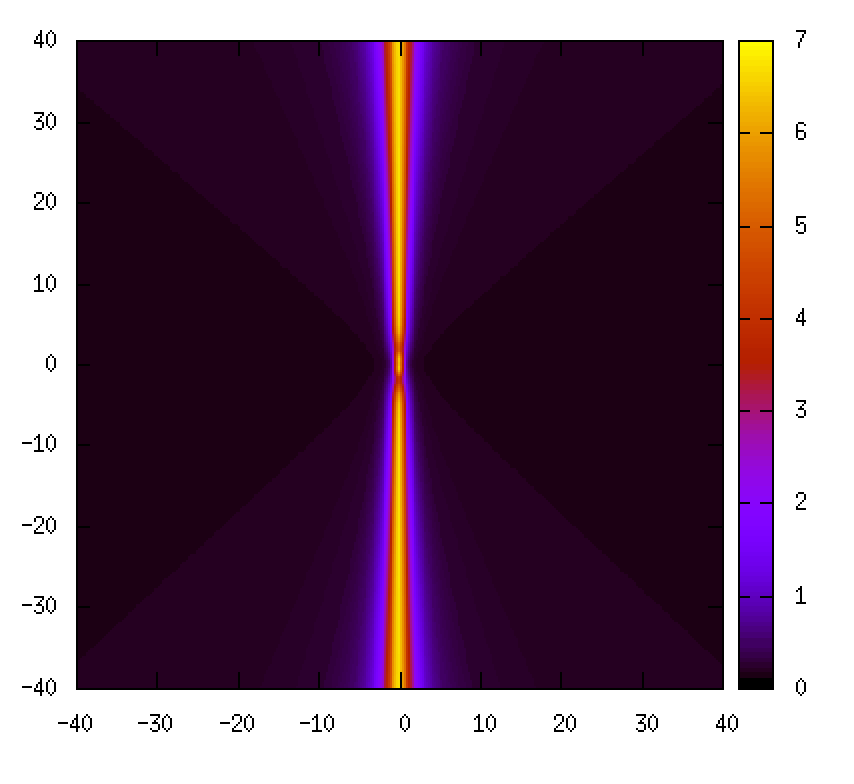}}
\subfigure[\, ]{\includegraphics[totalheight=4.0cm]{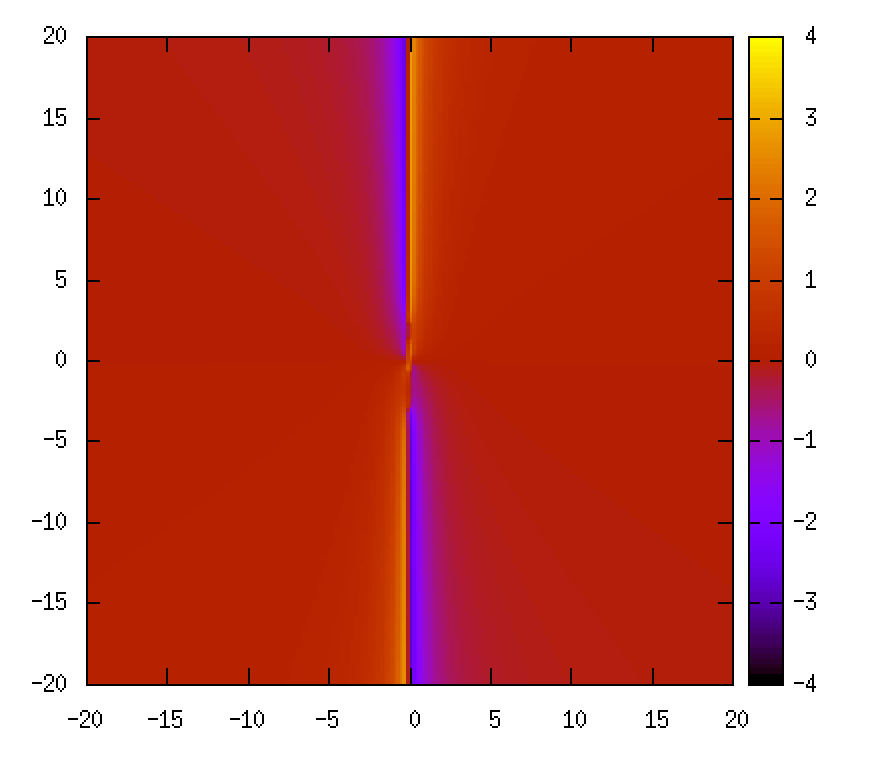}}
\caption{Figures of two domain walls attached to one Abelian axion string
before the decay for $\mu=1.4,~g=1,~\eta = 0.5, ~\lambda_1 = 2.0$ 
and $\lambda_2 = 0.25 $.
The string exists at the center in each figures.
They show: (a) $2 \Tr(({\Phi}^2 )^\dagger\Phi^2)$, (b) the energy density, and
(c) the phase $2\alpha$.
It is noted that the $\alpha$ changes by $\pi$ at crossing the domain wall.}
\label{axion_dw3}
\end{figure}

First let us consider an Abelian axion strings. 
Below we show that 
a single Abelian axion string is attatched by two domain walls.
To understand a situation in the presence of the walls attached to the Abelian axion string, 
we shall consider 
an approximate ansatz of the string at a large distance as
\begin{eqnarray}
\Phi \sim 2\eta e^{i\alpha}\tau^1, ~~~ A_\mu = 0.
\end{eqnarray}
Substituting the above ansatz into Eq.~(\ref{HamiltonianDW}), we find
\begin{eqnarray}
\label{axiondwpot}
\mathcal{H} \sim 2 \eta^2\l[ \l(\p_i\alpha\r)^2 + \mu\l(1 - \cos2\alpha\r)\r].
\end{eqnarray}
The potential in the above Hamiltonian is nothing but the potential of Eq.~(\ref{Vaxion}) wth $\alpha =a/2\eta$.
As noted already,
$\alpha$ sweeps full circle ($0\le \alpha \le 2\pi$) around the axion string,
however, the potential shows that
system have two vacua at $\alpha =0$ or $\pi$ and this would create two domain walls attached to the Abelian axion string: $N_{\rm DW}=2$. 
$\alpha$ would be almost zero 
or $\pi$
everywhere
however changes at the place where domain walls are created. 
In other words, there are two different domain walls:
One is connecting the vacuum at $\alpha=0$ and that at $\alpha=\pi$. 
Another is doing the vacuum at $\alpha=\pi$ and that at $\alpha=2\pi \equiv 0$.
We call them DW1 and DW2 respectively.
The whole configuration can be regarded as 
a junction of these two domain walls (DW1 and DW2), whose junction line is nothing but an Abelian axion string.
Fig.~\ref{axion_dw3} shows 
a full numerical simulation of such a configuration, obtained by the relaxation method.
We used larger lattice of $700\times 700$ points with lattice spacing $0.2$. 
We have taken $g=1,~\mu = 1.4,~\eta = 0.5, ~\lambda_1 = 2.0$ 
and $\lambda_2 = 0.25 $ for our computation of domain wall-string composites. 
Here, we put a large friction around a string to prevent this configuration to decay,  as described below.
%
%
\subsection{Alice axion string-domain wall composite}
In this subsection we discuss the formation of domain walls in the presence of two 
Alice strings produced via the decay of the parent Abelian axion string. 
To understand behavior of the walls, 
we similarly
start with field configurations at a large distance as
\begin{eqnarray}
\Phi \sim \eta\left(
\begin{array}{ccc}
0  &  e^{i\alpha}     \\
1  &  0  \\
\end{array}
\right),  \qquad
A_i \sim -\frac{1}{2g}\frac{\epsilon_{ij} x^j}{r^2} \tau^3 ~(i,j=1,2),~
A_0=A_3 =0.
\end{eqnarray}
With this ansatz, the static Hamiltonian density in Eq.~(\ref{HamiltonianDW})
reads
\begin{eqnarray}
\label{alivedwpot}
\mathcal{H} \sim \eta^2 \left[ \half (\p_i\alpha)^2 + 2 \mu \l(1 - \cos\alpha\r) \right].
\end{eqnarray}
Here we may check the difference from Eq.~(\ref{axiondwpot}). 
In this case, the vacuum is still at $\alpha = 0$ (or $2\pi$) whereas
the field range is given by $0\le \alpha \le 2\pi$.
As a result, there will exist only one domain wall attached to one Alice axion string. 
This is a similar to the vacuum with $N_{\rm DW} =1$.
The model is also identical to the sine-Gordon model in two dimension and a domain wall solution along the $x$-axis interpolating between the two vacua can be written as
$\alpha = 4 \tan^{-1} e^{\pm \sqrt{2\mu} x}$.

\begin{figure}[htbp]
\centering
\subfigure[\, ]{\includegraphics[totalheight=4.0cm]{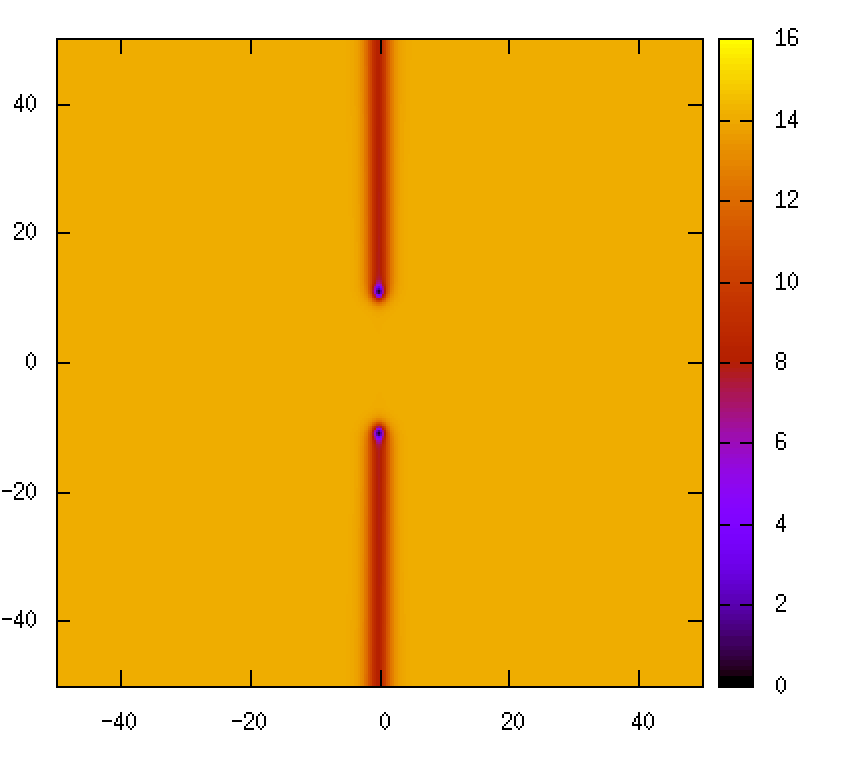}}
\subfigure[\, ]{\includegraphics[totalheight=4.0cm]{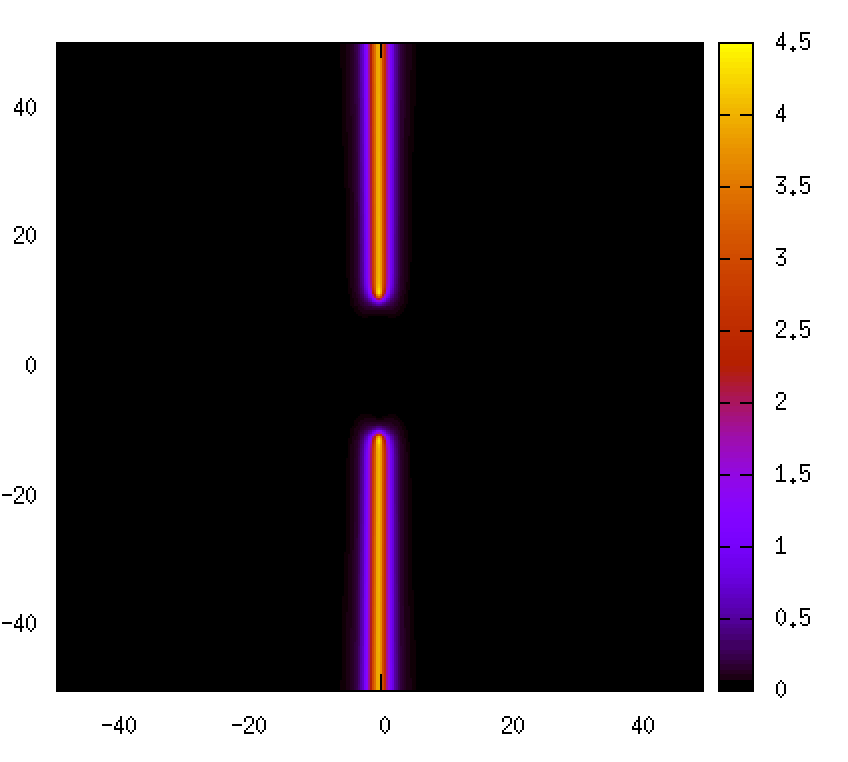}}
\subfigure[\, ]{\includegraphics[totalheight=4.0cm]{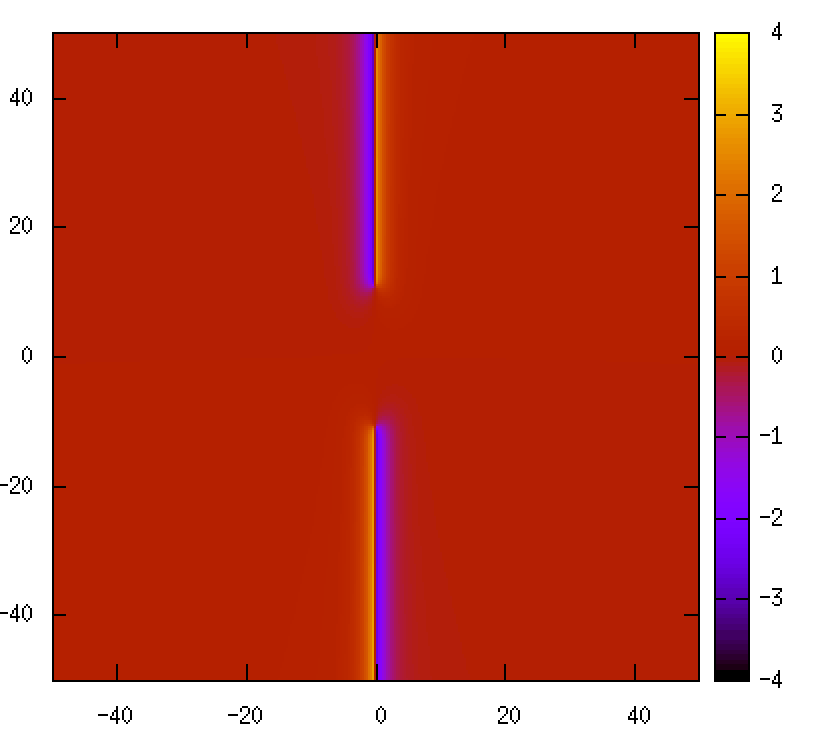}}
\subfigure[\, ]{\includegraphics[totalheight=4.0cm]{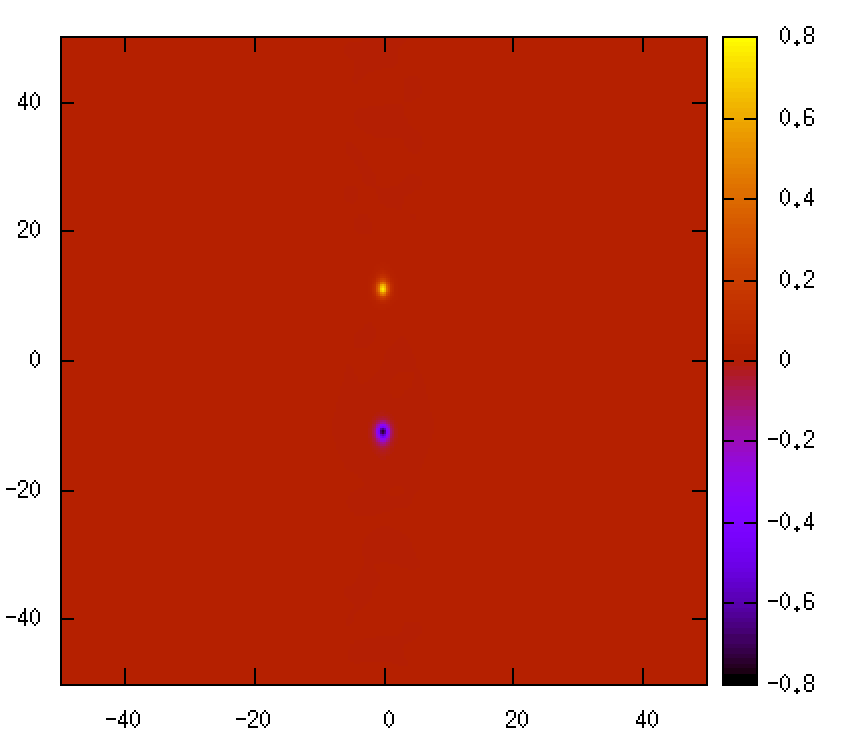}}
\subfigure[\, ]{\includegraphics[totalheight=4.0cm]{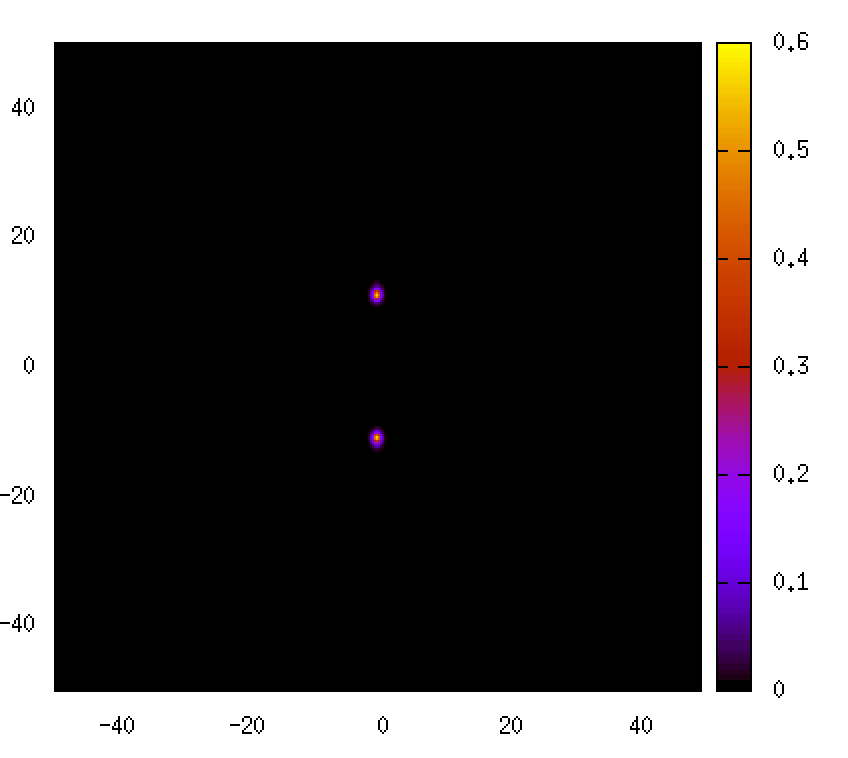}}
\caption{
Figures of two domain walls attached to two each Alice axion strings
after the decay of the parent Abelian axion string 
for $\mu=1.4~g=1~\eta = 0.5, ~\lambda_1 = 2.0$ and $\lambda_2 = 0.25 $.
They show: (a) $2 \Tr(({\Phi}^2 )^\dagger\Phi^2)$, 
(b) energy density,
(c) the phase $\alpha $,
(d) non-Abelian magnetic field $F_{12}^3(x,y)$, 
(e) the magnitude of the magnetic fields $\sum_\a {F^\a}^2_{12}$.
It is noted that $\alpha$ changes by $2 \pi$ at crossing the domain wall.
}
\label{Scalar}
\end{figure}

For $\mu\ne 0$ in computation, 
when an Abelian axion string attached by two domain walls 
is initially created and decays into two Alice axion strings, 
each domain walls remains attached to Alice strings as shown in the Fig.~\ref{Scalar}.
For numerical calculations, the relaxation method is used and the same parameters are
chosen as before. 

\begin{figure}[htbp]
\centering
{\includegraphics[totalheight=5.0cm]{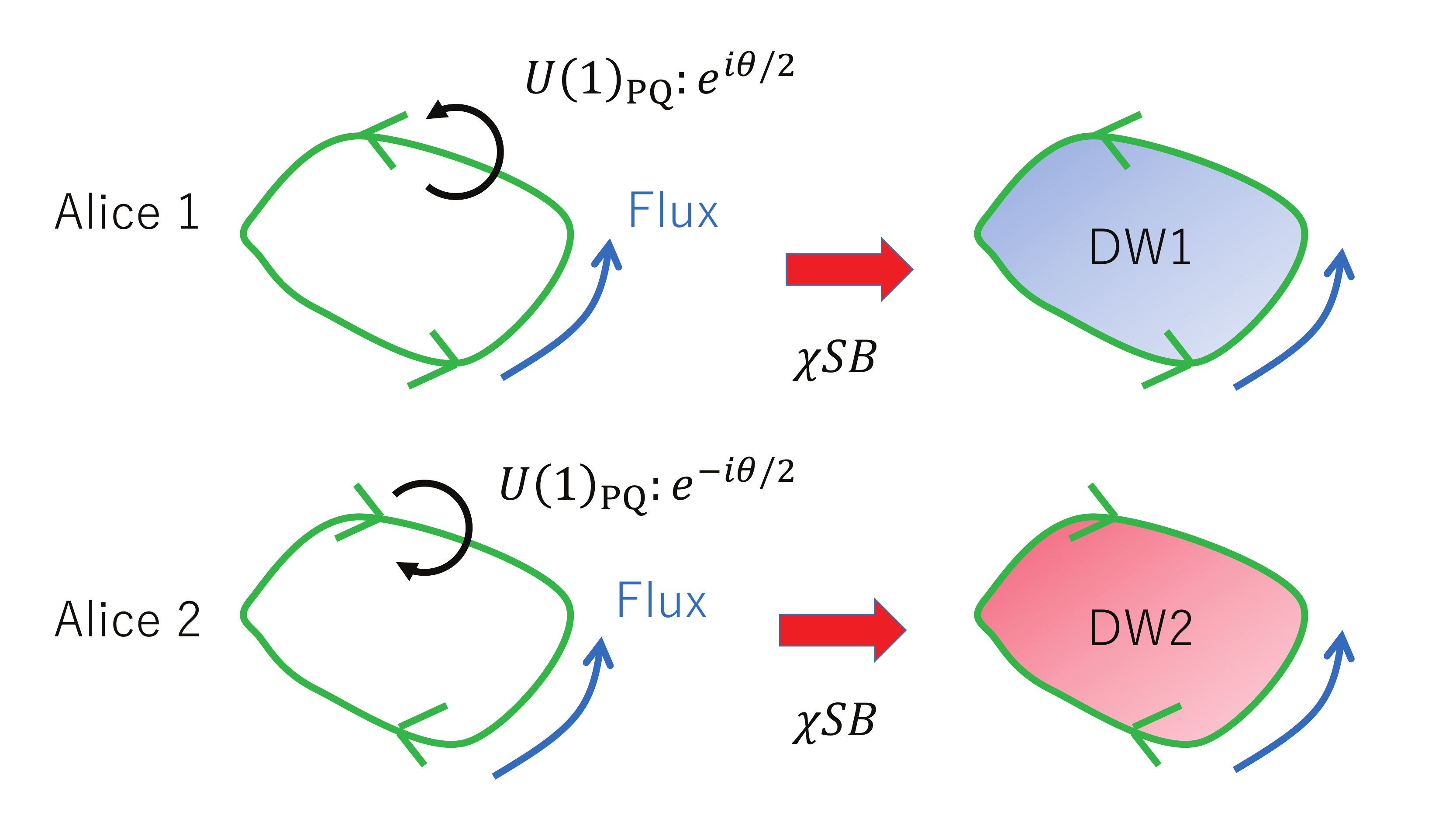}}
\caption{
A schematic figure for Alice strings attached by domain walls. 
There exist also two kinds of Alice string depending on the orientation and the flux.
At the chiral phase transition, 
DW1 is attached to Alice1 whereas DW2 is attached to Alice2.
}
\label{Alice_DW}
\end{figure}

There exist two kinds of the Alice axion strings when
one focuses on Eq.~(\ref{aliceaxion}).
One has the non-Abelian flux parallel 
to the orientation defined by the $U(1)_{\rm PQ}$: 
$e^{i\theta/2}\Omega = e^{i\theta/2} e^{i\theta \tau^3/2}$, where
$e^{i\theta/2} \in U(1)_{\rm PQ}$.
Another has the flux opposite
to the orientation: 
$e^{i\theta/2}\Omega = e^{i\theta/2} e^{-i\theta \tau^3/2}$
(with a modulus parameter of $\pi$).
We call these strings Alice1 and Alice2 respectively.
To flip the sign simultaneously by $\theta \to -\theta$ 
gives the same configuration because this is 
to see the same string in different ways: 
the string seen from a positive $z$ coordinate
or from a negative $z$ coordinate. So, 
Alice2 with $e^{i\theta/2} e^{-i\theta \tau^3/2}$ is physically equivalent to 
that with $e^{-i\theta/2} e^{i\theta \tau^3/2}$.
Factors of $e^{\pm i\theta/2}$ in $U(1)_{\rm PQ}$ 
(with the same $\Omega =e^{i\theta \tau^3/2}$)
imply the way to approach domain walls in the axion space.
After one rotation around the Alice1 with $0 \leq \theta/2 \leq \pi$,
one meets DW1.
Further, with $\Omega(2\pi) = e^{i \pi \tau^3} \in {\mathbb Z}_2 \subset H$,
the vacuum is smoothly connected.
At the chiral phase transition, like in $N_{\rm DW} =1$ case, 
a single Alice1 is attached by a single DW1, 
whereas a single Alice2 is attached by a single DW2. See Fig.~\ref{Alice_DW}.

\begin{figure}[htbp]
\centering
\subfigure[\, ]{\includegraphics[totalheight=4.0cm]{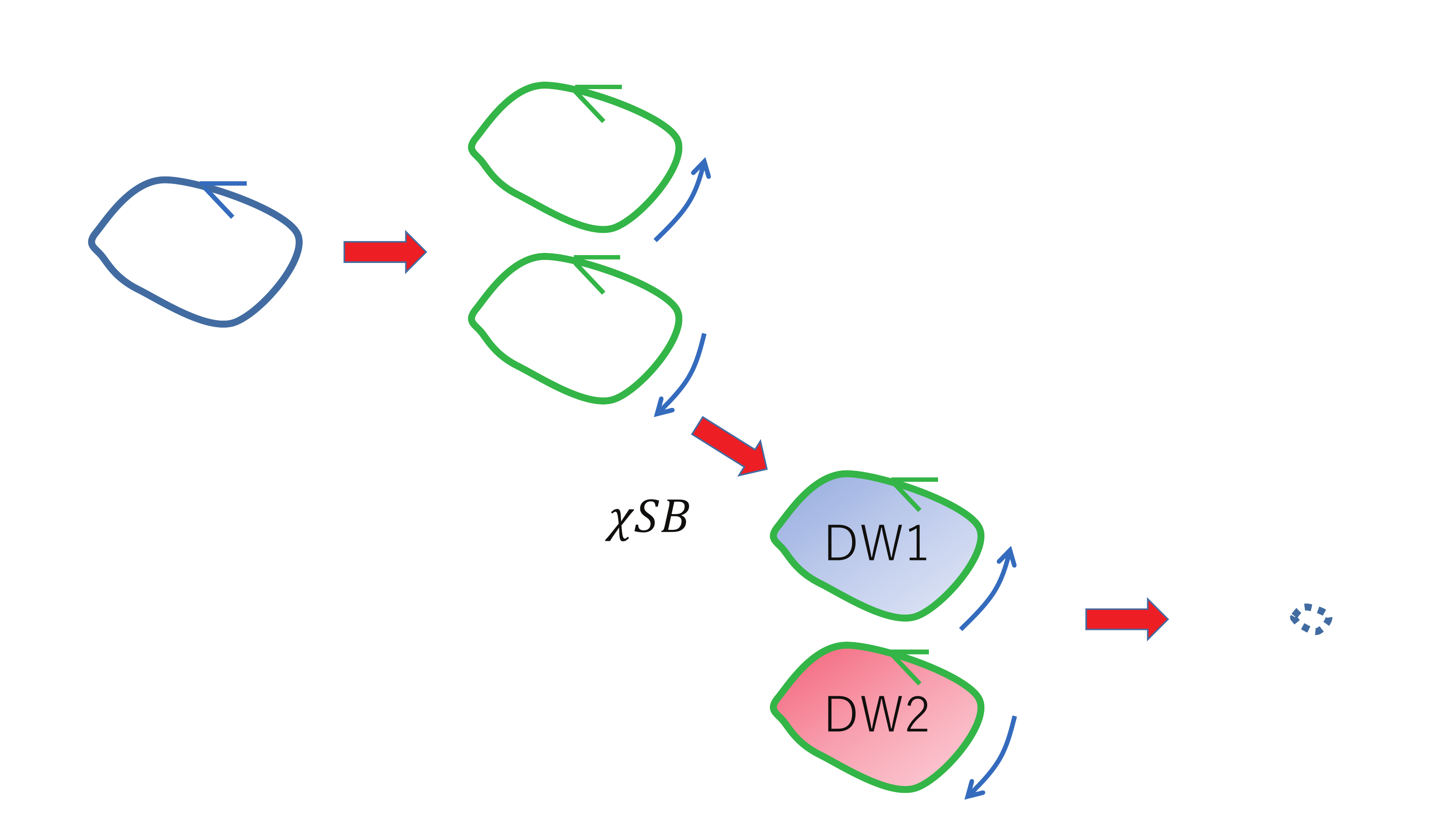}}
\subfigure[\, ]{\includegraphics[totalheight=4.0cm]{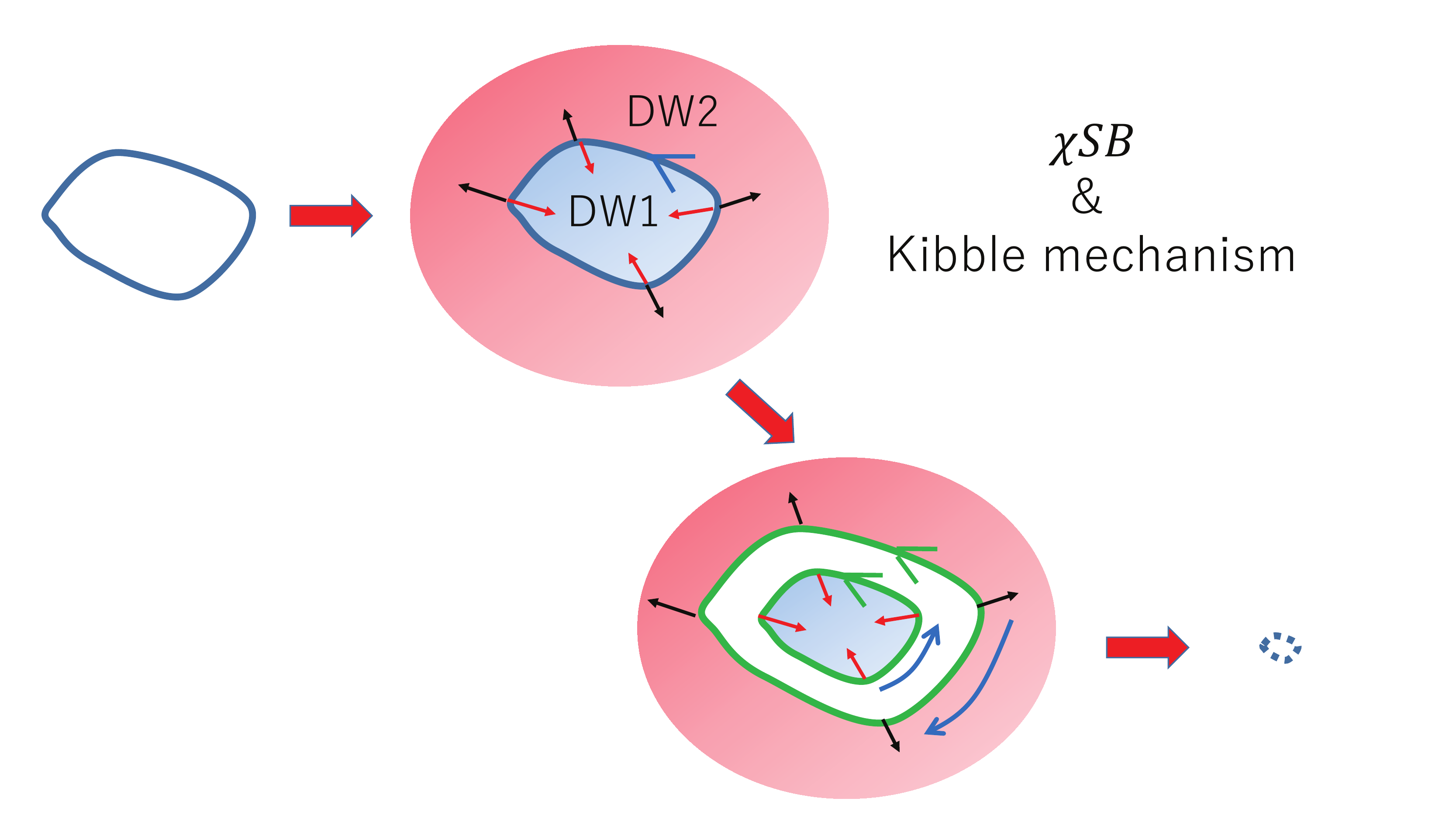}}
\caption{
Schematic figures for decay of Abelian axion string into two Alice strings attached by domain walls. 
(a) An Abelian axion string can quickly decay into two Alice strings in the early universe.
(b) Even if an Abelian string survives at the chiral phase transition, 
it can be split into two Alice strings attached by domain walls in the presence of domain wall tension.
In both cases, each of the walls shrinks to a point by the tension finally.
}
\label{Alice_DW2}
\end{figure}

In the actual cosmological history,
once an Abelian string is created after the PQ symmetry breaking, 
it will quickly decay into two Alice strings in the presence of the repulsive force
between them. Then domain walls, which are created
after the chiral symmetry breaking,
attach to Alice strings, and they shrink to a point owing to the wall tension
like in the $N_{\rm DW} =1$ case.
Even if an Abelian sting attached by two domain walls is created at the chiral phase transition by the Kibble-Zurek mechanism, 
the Abelian string can be split into two Alice strings in the presence of the wall tension.
Then, one Alice string is attached by one wall
and each of the walls similarly shrinks to a point by the tension. See Fig.~\ref{Alice_DW2}.
%
\section{Conclusion and discussions}
\label{conclusion}
Axion is an attractive candidate of dark matter, while
stable networks composed by stable strings and walls may be created 
after the breakdown of $U(1)_{\rm PQ}$ and 
${\mathbb Z}_{N_{\rm DW}}$ embedded in $U(1)_{\rm PQ}$.
They can cause cosmological disasters since the energy density of them 
can finally dominate the that of the universe.
The Lazarides-Shafi mechanism is 
one of solutions to
the domain wall problem. 
We have studied this mechanism in detail 
based on a recently proposed model \cite{Sato:2018nqy},
showing dynamics of axion strings and walls.
Even if Abelian axion strings are formed in the early universe, each of them is split into multiple Alice axion strings 
due to a repulsive force between the Alice strings even without domain wall.
When domain walls are formed as the universe cools down, a single Alice string is attached by a single wall 
because a vacuum is connected by a non-Abelian rotation without changing energy. 
Such walls do not form stable networks since they collapse by the tension of the walls, emitting axions.
Even if domain walls attached to the Abelian axion strings is created 
by the Kibble-Zurek mechanism at the chiral phase transition, 
the Abelian string can be split into Alice strings and 
one domain wall is attached to one Alice axion strings.
Such walls can shrink to a point owing to the wall tension
like in $N_{\rm DW}=1$ case.

Several discussions are addressed here.
The model of Ref.~\cite{Sato:2018nqy} was also proposed as a model for 
a monopole dark matter.
A monopole in the conventional Alice theory ($SO(3)$ gauge theory with scalar fields of the fiveplet) admitting Alice strings was studied in 
Refs.~\cite{Shankar:1976un,Bais:2002ae,Striet:2003na,Benson:2004ue}. 
In particular, a monopole is not spherical and 
decays into a twisted Alice ring depending on choice of parameters
\cite{Bais:2002ae,Striet:2003na}.
It would be interesting to study 
if the same would happen in our case. 
In fact, it is known that 
a global analog (global monopole) shows this property 
\cite{Ruostekoski:2003qx}.
A dyon with an electric $U(1)_H$ charge may be realized as a vorton, 
namely a persistent electric current flows along a ring.
Dyons can be dark matter if their (mini) electric charge, which may be obtained via
a kinetic mixing between $U(1)_H$ and electromagnetism,
is below experimental bounds.
It is also worth to point out that 
the conventional monopole charge of $\pi_2$ is not well defined in the presence 
of an Alice string, because 
a monopole becomes an anti-monopole 
when it encircles around an Alice string, 
as a dual of electric charge encircling around the Alice string. 
Instead of using the usual homotopy group $\pi_2$, 
a monopole charge must be defined in terms of the Abe homotopy 
\cite{Kobayashi:2011xb}. 

There can exist infinitely long (Abelian) strings produced after the PQ breaking.
In such cases, the scaling solution is found to be violated
by a logarithmic growth of the string scaling 
parameter in time \cite{Gorghetto:2018myk,Kawasaki:2018bzv}.
It might be hard for wide walls attached to long strings to shrink to a point,
hence simulations for them may also be altered.
When two Abelian cosmic strings collide, they reconnect each other,
which is important process for cosmic strings to reduce their number. 
Alice strings have $U(1)$ moduli corresponding to fluxes, 
and so it is unclear if they can reconnect.
As this regards, two non-Abelian strings with non-Abelian fluxes 
were shown to always reconnect \cite{Eto:2006db}, 
and so it would be true for Alice strings.
Further, 
a nature of reconnection among Alice strings may be different
from that among Abelian strings
due to a force existing among Alice strings, so
the number of long Alice strings could differ from that of long Abelian string.
The number of long Alice strings would be significant to a solution to the domain wall problem.
In any cases, the axion abundance needs to be correctly estimated
and may be modified from Eq.~(\ref{Omegaa}).
Thus, an allowed region for the axion decay constant may be altered.
If there is no allowed region,
the PQ symmetry breaking might be required to take place before or during inflation,
and constraint on insocurvature may be important then. 

In future observations, gravitational waves produced by the 
decay of strings and walls may be detected, depending on axion model \cite{Saikawa:2017hiv,Higaki:2016jjh}.
That can be an important signal to verify the presence of axion dark matter
produced by the topological objects.

If topological objects appear in dark matter models,
it is necessary to study the nature of the objects in detail,
for precise estimation of dark matter.

\section*{Acknowledgments}
MN would like to thank Naoyuki Takeda for his lecture of axion cosmology.
This work is supported by the Ministry of Education, Culture, Sports, Science and Technology (MEXT)-Supported Program for the Strategic Research Foundation at Private Universities ``Topological Science" (Grant No. S1511006). 
C.~C. acknowledges support as an International Research Fellow of the Japan Society for the Promotion of Science (JSPS) (Grant No: 16F16322). 
This work is also supported in part by 
JSPS Grant-in-Aid for Scientific Research (KAKENHI Grant No. 16H03984 (M.~N.), No.~18H01217 (M.~N.)), and also by MEXT KAKENHI Grant-in-Aid for Scientific Research on Innovative Areas ``Topological Materials Science'' No.~15H05855 (M.~N.).

\end{document}